\documentclass[11pt,Times]{article}
\usepackage[sort&compress,numbers]{natbib}
\usepackage{float}

\usepackage{amsmath, amsfonts, amsthm}
\usepackage{enumerate}
\usepackage{geometry} 
\usepackage{graphicx,subfigure}
\usepackage{color}
\usepackage{hyperref}
\usepackage{setspace}

\usepackage{bm}

\usepackage{geometry}
\def\cite{\citep}

\def\ii{{(i)}}
\def\kk{{(k)}}
\def\Theta{\xi}
\def\g{\gamma}
\def\lam{t}
\def\l{\lambda}
\def\dir{.}

\begin{document}


 \title{Detecting the structure of haplotypes, local ancestry and \\excessive local European ancestry in Mexicans}

\author{Yongtao Guan \protect\\ Baylor College of Medicine}
\date{}
\maketitle


\begin{abstract}
We present a two-layer hidden Markov model to detect structure of haplotypes for unrelated individuals. This allows modeling two scales of linkage disequilibrium (one within a group of haplotypes and one between groups), thereby taking advantage of rich haplotype information to infer local ancestry for admixed individuals. 
Our method outperforms competing state-of-art methods, particularly for regions of small ancestral track lengths. 
Applying our method to Mexican samples in HapMap3,  we found five coding regions, ranging from $0.3 -1.3$ megabase (Mb) in lengths, that exhibit excessive European ancestry (average dosage $> 1.6$). A particular interesting region of $1.1$Mb (with average dosage $1.95$) locates on Chromosome 2p23 that harbors two genes, PXDN and MYT1L, both of which are associated with autism and schizophrenia. In light of the low prevalence of autism in Hispanics, this region warrants special attention. We confirmed our findings using Mexican samples from the $1000$ genomes project.  A software package implementing methods described in the paper is freely available at \url{http://bcm.edu/cnrc/mcmcmc}. 

\end{abstract}


\section{Introduction}

Haplotype variation is central to statistical and population genetics.
Studies have revealed that considerable sharing of haplotypes exists across populations~\cite{conrad.etal.06}, as well as significant variation among populations~\cite{liu.etal.04}.  Polymorphic markers are linked on a haplotype; thus, differences in haplotype abundance cause linkage disequilibrium (LD, the nonindependence of marginal allele frequencies)  between markers.  Therefore, modeling LD helps to understand haplotype variations. 
Many statistical models exist to model LD; however, a model to detect the structure of haplotypes is missing.  

The most elegant model for LD is the coalescent with recombination~\cite{kingman.82,hudson.83}, or ancestral recombination graph (ARG). However, despite successful efforts on small-scale datasets~\cite{wang.rannala.09}, ARG remains notoriously hard to compute. 
Considerable efforts have been made to  approximate ARG  to allow computation on a large scale~\cite{stephens.donnelly.00, fearnhead.donnelly.02, li.stephens.03,fastphase,paul.song.10}. Among them, the most successful is the PAC model of \citet{li.stephens.03}, which models a new haplotype as an imperfect mosaic of observed haplotypes to produce a conditional likelihood; the joint likelihood of all haplotypes is then approximated by the product of those conditionals. 
 Along these lines, \citet{paul.song.10} made formal derivation using diffusion approximation. 
A somewhat related approach is the clustering model~\cite{fastphase}, which summarizes observed haplotypes into a small number of (ancestral) haplotypes and models the observed haplotypes as imperfect mosaics of those haplotypes. 

These models assume that haplotypes are sampled from a single source population and become ineffective when haplotypes are admixed.  Admixed haplotypes have two scales of LD: the admixture LD that formed between alleles in different source populations and typically spans a few to tens of centimorgans (cM)~\cite{mald}; and the LD between alleles within each source population that typically spans a few tenths of a cM.   
The HAPMIX~\cite{hapmix} model is among the first to model LD of admixed individuals, extending the PAC model to two source populations. This model is very effective for inferring local ancestry for two-way admixtures (e.g.,~African Americans), but it is not yet applicable to three-way admixtures such as Latinos (in principle, however, HAPMIX should work with three-way admixtures).  Two recent examples of progress include LAMP-LD~\cite{lampld} and MULTIMIX~\cite{multimix}, both of which achieve similar performance with HAPMIX in inferring the local ancestry of two-way admixtures, and can deal with three-way admixtures. However, HAPMIX and LAMP-LD both require haplotypes from source populations, and LAMP-LD and MULTIMIX both assume that ancestries are fixed within a window of loci and only switch between windows. These methods often perform well for recent admixture but underperform for distant admixture, which implies limited ability to detect local ancestries of short track lengths. In addition, distantly admixed individuals, such as Uyghurs whose admixture occurred more than $100$ generations ago, are valuable for disease association~\cite{xu.jin.08} and human genetic landscape studies~\cite{huili.etal.09}.   

A different perspective of two scales of LD in admixture is  \emph{structure on local haplotypes}. Taking two-way admixture as an example, haplotypes from two source populations are separated into two groups, and a new haplotype is assigned probabilistically to a group based on its similarity with haplotypes in the group. Grouping source haplotypes is equivalent to putting a structure on haplotypes.  In fact, the structure of local haplotypes is an ubiquitous phenomenon in genetic data, and the admixture is just a more apparent example.  Even among individuals sampled from a single source population, a set of local haplotypes might be enriched in one subset of individuals and a different set of local haplotypes enriched in another. For example, individuals of European descents may be separated according to whether they have different two-digits human leukocyte antigen  (HLA)-A allele classes. Compared to differences in local ancestry, the difference in two-digits HLA allele classes is more subtle. However, from the perspective of statistical modeling, these two scenarios are the same -- both require detecting the structure of local haplotypes based on their similarities.  
None of the current methods is designed to handle this more delicate scenario. 

In this study, we present a novel two-layer hidden Markov model (HMM) designed to learn the structure of local haplotypes. The new model uses two layers of latent clusters.  In each layer, clusters are labeled to represent ancestry alleles, and multiple clusters of the same label over adjacent loci represent an ancestral haplotype.  
In a nonrecombined region, the upper layer aims to capture structure near the root of a coalescent tree, whereas the lower layer aims to capture haplotype variation near the tip.  
Recombination is approximated by cluster switching within each layer. 
The lower layer clusters are fuzzy mosaics of the upper layer clusters, and haplotypes in the observed data are fuzzy mosaics of the lower layer clusters. The fuzzy mosaic represents the mode of inheritance for haplotypes: mosaic implies historic recombinations and fuzziness implies mutations and uncertainty of inheritance. 
Existing cluster-based models use single-layer clusters. For example, fastPHASE~\cite{fastphase} and Beagle~\cite{beagle} use, equivalently, the lower layer clusters to model ancestral haplotypes; and the STRUCTURE~\cite{structure} equivalently use the upper-layer clusters to model ancestry populations.  Although seemingly incremental, the two-layer model has an attractive feature that is not available in a single-layer model -- detecting \emph{structure of haplotypes}.  The upper-layer clusters represent different groups (populations) and the lower-layer clusters represent group-specific haplotypes. This allows us to 1) summarize local haplotypes into different groups and 2) assign a local haplotype probabilistically into groups, which are two main ingredients for local ancestry inference. 

Local ancestries of admixed individuals provide important information for disease association mapping~\cite{mald} and demographic history~\cite{hua.11}. It is an important subject that has attracted much recent attention~\cite{ancestrymap,tang.etal.06,hapaa,hapmix,lampld,multimix}.
One way to infer local ancestry is to use ancestry informative markers (AIMs) -- loci in which the allele frequencies have large differences among populations~\cite{aim}. Inferences based on AIMs usually have low resolutions because AIMs are relatively scarce. On the other hand, haplotypes provide richer information that is complementary to the AIMs. Taking an extreme example, if population $1$ has $50\%$ A-T and $50\%$ T-A haplotypes whereas population $2$ has $50\%$ A-A and $50\%$ T-T haplotypes, there would be no difference in the marginal allele frequencies between two populations. However, the  two-marker haplotypes are very informative.
The two-layer model uses local haplotypes in source populations to define population features for each small genomic region, and based on which admixed haplotypes are assigned  probabilistically to different populations. These genomic regions are not prespecified; instead, they are learned from data. Compared to methods that group markers in windows and only allow ancestral switches between windows~\cite{lampld,multimix}, our method has better performance because prespecified windows may conflict with actual ancestral switches.  

\section{Results}
The model and its computation are described in details in the Methods section. Briefly,  each haplotype associates with latent states of $S$ upper clusters and $K$ lower clusters at each marker. We fit the model using the Expectation Maximization (EM) and compute the cluster dosages conditional on data to infer local ancestry.  Let $\g$ denotes the admixture generation, and let $\l=100/\g$ denotes the average ancestral track length (in cM). 

\begin{figure}[t]
\centering
\includegraphics[width=14cm]{\dir/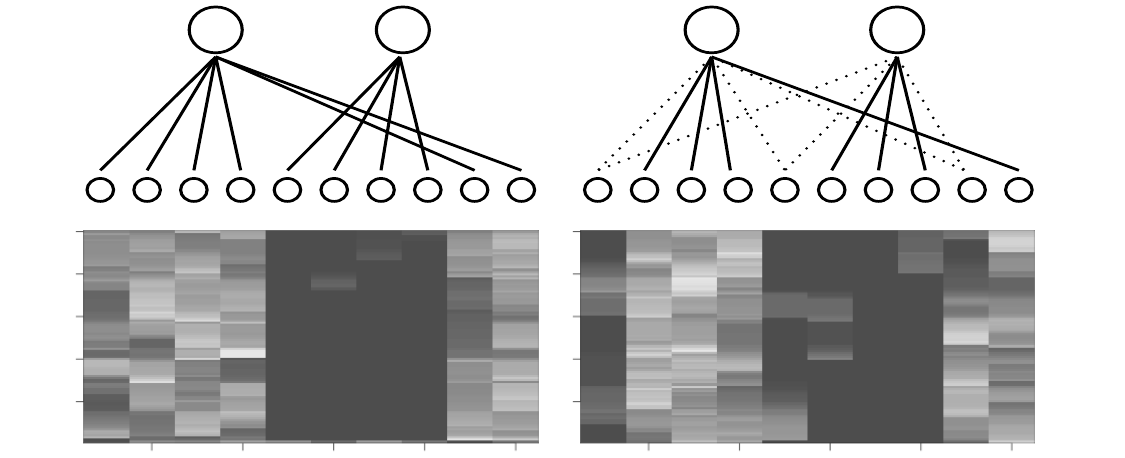}
\caption {Structure of haplotypes. Each row denotes a SNP, and each column denotes a lower-layer haplotype in our model. We choose two typical regions that each contain $100$ SNPs. The plot shows the lower clusters dosage conditional on one upper cluster (conditional dosage).  Brighter pixels indicate higher dosages.  The solid edge in the diagram indicates that the conditional dosage of a haplotype is either $>80\%$ (connecting to the left upper cluster) or $<20\%$ (connecting to the right upper cluster). Otherwise, we draw a dashed line indicating edge uncertainty.}
\label{fig:structure}
\end{figure}

\subsection{Structure of haplotypes}
The two-layer model can detect the structure of haplotypes. To illustrate this, we took the Chromosome $2$ of unrelated CEU and YRI individuals from HapMap2~\cite{hapmap} and fit the two-layer model with $S=2, K=10$ and $\g=100$, ignoring their population labels.  Then, we computed the lower cluster dosage conditional on an upper-layer cluster (conditional dosage) for each individual and averaged the conditional dosages over all individuals. The conditional dosages for two typical regions ($100$ SNPs each) were plotted in Figure~\ref{fig:structure}.  In one region, the lower clusters are split rather cleanly (but not evenly) between two upper-layer clusters;  in the other, the lower-layer clusters are split but less cleanly with some lower clusters shared between two upper clusters. 
This example illustrates that the two-layer model can indeed detect the structure of haplotypes. Moreover, Figure~\ref{fig:structure} demonstrates that some local haplotypes are population-specific whereas others are shared between populations. This \emph{local haplotype sharing} is an intrinsic feature of genetic data~\cite{conrad.etal.06}. The two-layer model can learn this feature, which is of particular importance in local ancestry inference.  
As a comparison, local haplotype sharing is not a natural part of the HAPMIX~\cite{hapmix} model, and a miscopy parameter is introduced and (somewhat) arbitrarily specified to adapt to the local haplotype sharing feature of the data.

\subsection{Local ancestry inference}\label{subsec:lai}

\begin{figure}[htp]
\centering
\includegraphics[width=14cm]{\dir/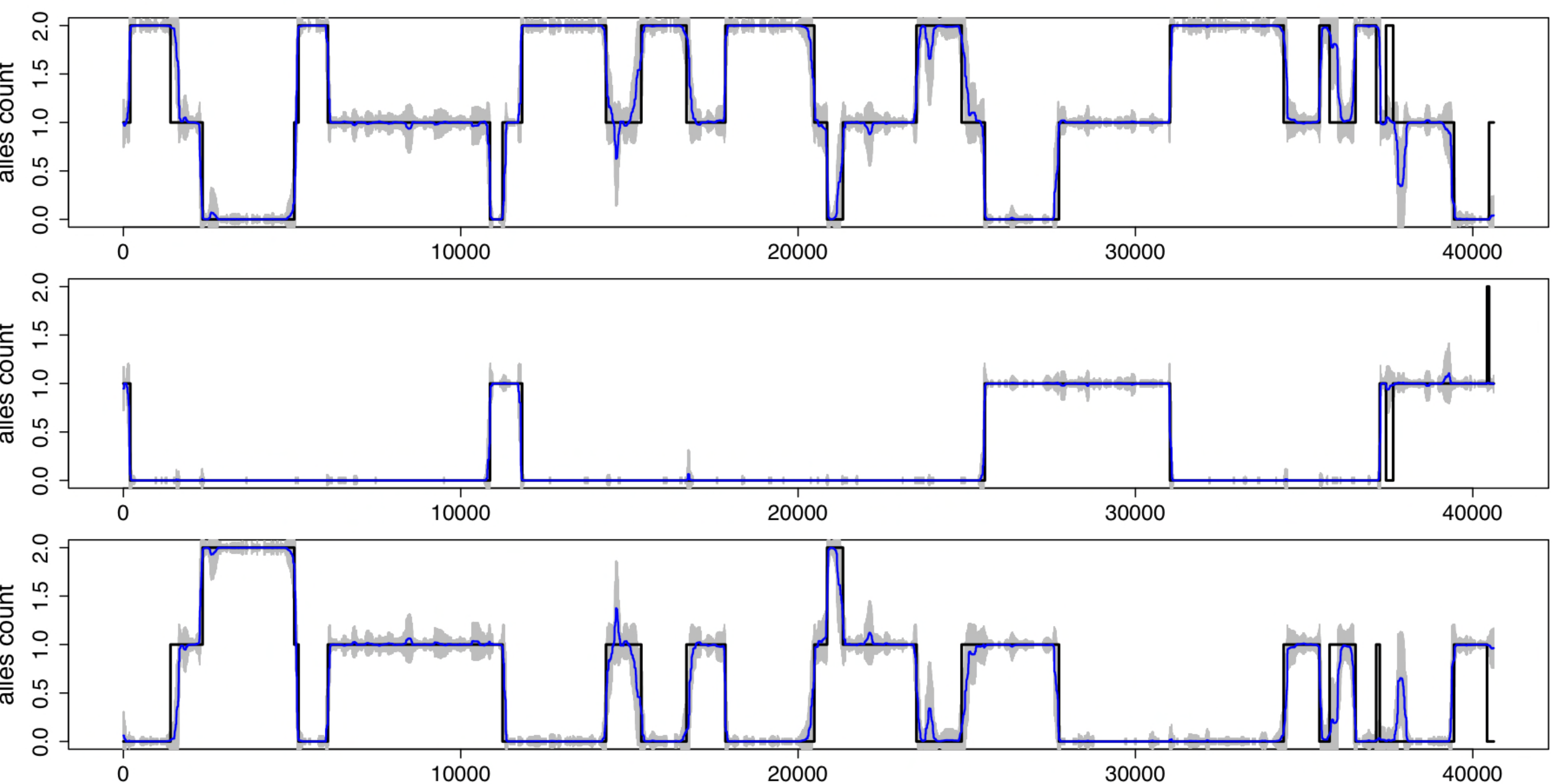}
\caption{Inference of local ancestry. The plot shows the results of a typical EM run for local ancestry inference of a three-way admixed individual. In each panel, x-axis denotes SNPs along the chromosome, and y-axis denotes ancestry allele dosage. The black lines in each panel are the actual values, and the blue lines are inferred mean dosages. The gray bars on top of blue lines reflect $\pm2$ standard deviations of the estimated mean dosages. At each locus, the y-values on lines of same color sum to $2$.}
\label{fig:3mix}
\end{figure}

We first illustrate that our method can achieve exceptional accuracy in local ancestry inference.  We simulated a three-way admixed individual ($\g=20$, Supplementary Note) and fit the two-layer model ($S=3, K=15$) using this individual and individuals from source populations, excluding haplotypes used to simulate the admixed individual.  Figure~\ref{fig:3mix} demonstrates the comparison between the true and inferred local ancestries.  The two-layer model can infer the local ancestry for a three-way admixed individual with exceptional accuracy. The loci to which inferred ancestral allele dosages do not match well with the truth tend to have large uncertainties.  This suggests that when combining results over multiple EM runs, the estimates may be weighted by their uncertainty, e.g.,~inverse of variance.  Note that for a diploid individual, our method can compute the probabilistic assignment to all possible pairs of ancestries at each marker, allowing us to quantify the mean and variance of the estimated ancestry dosages.  The admixture proportions were also accurately inferred (Supplementary Fig.~S1). 

\medskip
{\bf Comparison with HAPMIX and LAMP-LD.}
Next, we compared our method with two state-of-art methods used in local ancestry inference: HAPMIX (for two-way admixture) and LAMP-LD (for three-way admixture).  We used two metrics in our comparison -- mean deviation and Pearson's correlation between the inferred and actual local ancestries for each simulated admixed individual.   
\begin{figure}[htp]
\centering
\includegraphics[width=7cm]{\dir/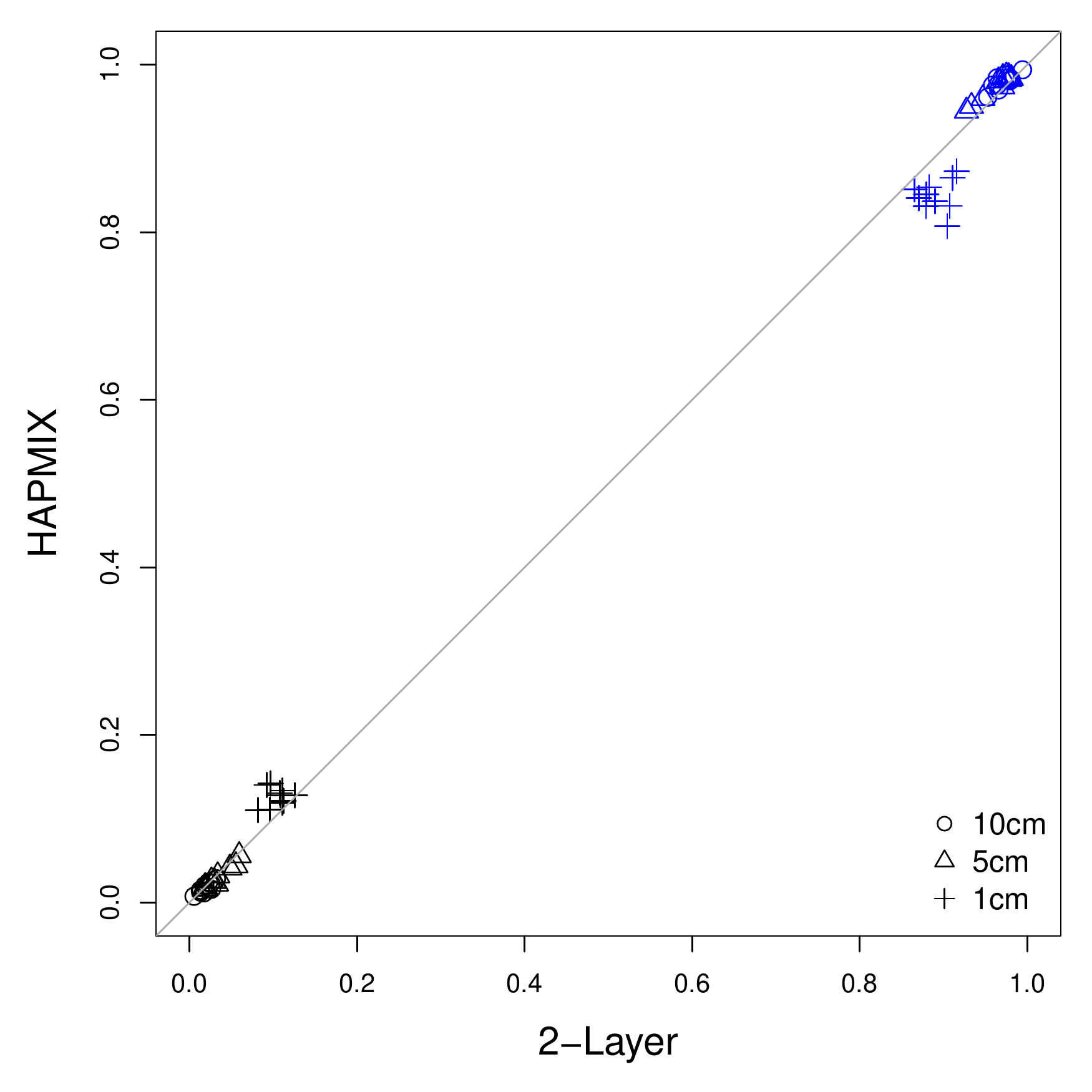}
\includegraphics[width=7cm]{\dir/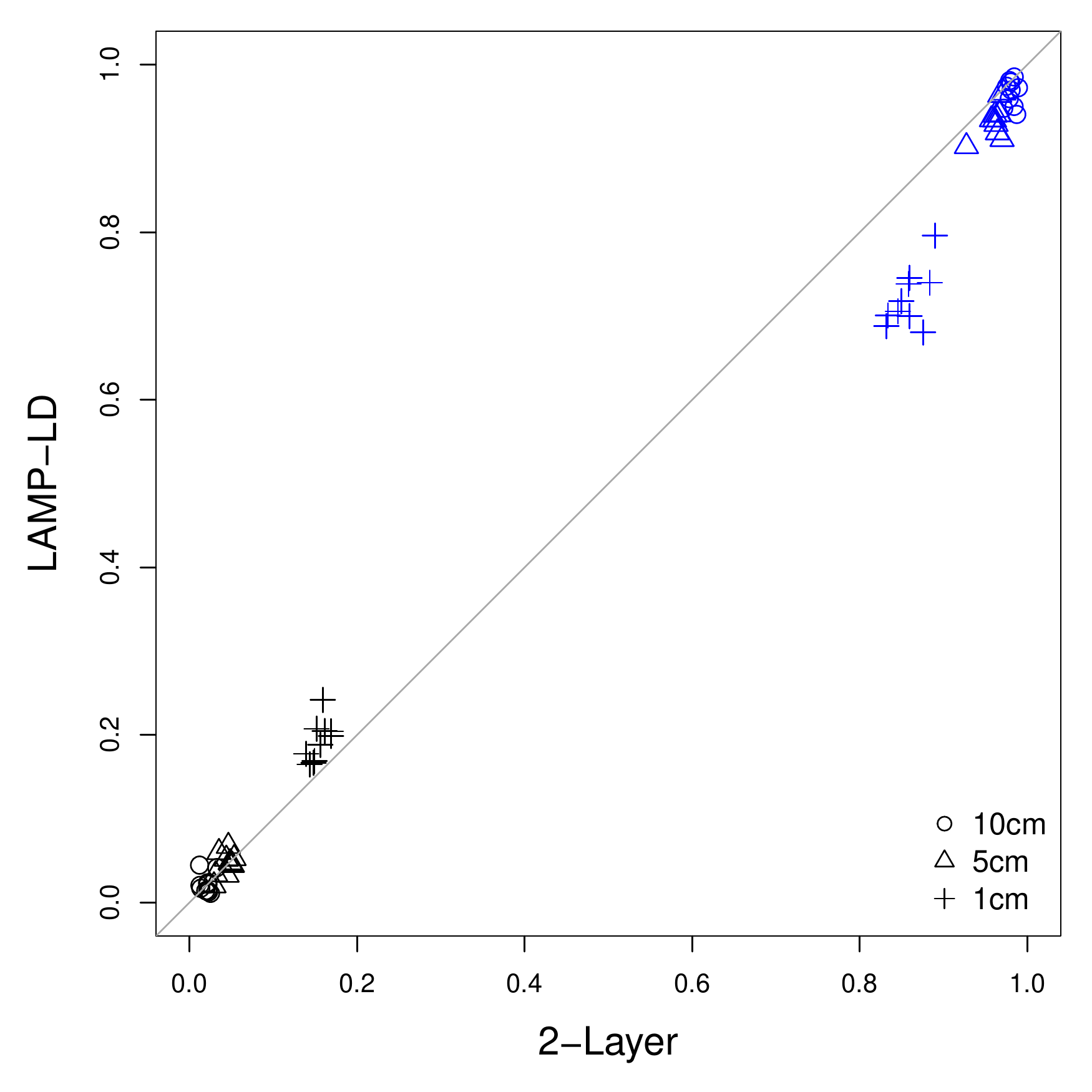}
\caption{Comparison with  HAPMIX and LAMP-LD. The left panel is for two-way admixture (HAPMIX) and the right panel is for three-way admixture (LAMP-LD). Results of our method are on the x-axis, and results of other methods are presented on the y-axis.    Each point in the plot represents a simulated admixed individual whose local ancestry track lengths are $10, 5$ and $1$cM as shown in the legend. Black points denote mean deviation, and blue points denote Pearson's correlation between the inferred and actual local ancestries.}
\label{fig:comparison:lamp1}
\end{figure}

For comparison with HAPMIX, we simulated three sets ($10$ individuals in each set) of two-way admixed individuals with $\g=10, 20$ and $100$ (corresponding to $\l=10, 5$ and $1$cM respectively). The difficulty in inferring local ancestry increases as the admixture generation increases. The results of our method were obtained with $S=2$ and $K=10$ and averaged over $10$ independent EM runs. The results of HAPMIX were obtained using its default parameters.    
 For easier problems ($\g=10,20$), when both methods perform well, HAPMIX performs slightly but not significantly better (two sample t-test $p=0.52, 0.63$ for deviation, and $p=0.20,0.09$ for correlation), whereas for harder problems ($\g=100$), our method outperforms HAPMIX ($p=5\times10^{-4}$ for deviation and $p=2\times10^{-5}$ for correlation) (Figure~\ref{fig:comparison:lamp1}). 
Our method has some practical advantages over HAPMIX: 1) it cleanly handles missing data, whereas HAPMIX does not allow missing data;  and 2) it can directly work with genotype data, whereas HAPMIX requires haplotypes from source populations.  When the phasing of individuals from source populations is  imperfect (e.~g.,~statistical phasing without the help of transmission), our method has an advantage. 

We compared our method with LAMP-LD for three-way admixed individuals.  Similar to the comparison with HAPMIX, we simulated three sets ($10$ individuals in each set) of three-way admixed individuals with $\g=10, 20$ and $100$. 
The results of our method were obtained with $S=3$ and $K=15$ and averaged over $10$ independent EM runs. The results of LAMP-LD were obtained with default parameters. 
Similar to the comparison with HAPMIX, for harder problems ($\g=100$), our method outperforms LAMP-LD (deviation $p=6\times10^{-4}$ and correlation $p=2\times10^{-8}$).   For easier problems ($\g=10$ or $20$), both methods perform similarly if measured by deviation ($p=0.69$ or $0.67$). There is a marked difference in performance if measure by Pearson's correlation -- our method outperforms LAMP-LD ($p=0.01$ for $\g=10$ and $p=3\times10^{-3}$ for $\g=20$) (Figure~\ref{fig:comparison:lamp1}). A closer look revealed that LAMP-LD tends to make more mistakes on small regions of a few hundred SNPs (Supplementary Fig.~S2).  We suspect that this has to do with grouping markers into windows, even though the recommended window size ($50-100$ SNPs~\cite{lampld}) is smaller than the size of often misidentified regions. In addition, LAMP-LD appears to be very certain everywhere, which can be misleading.   

\subsection{Excessive local European ancestry in Mexicans}
We applied our method to infer the local ancestries of Mexican samples in both HapMap3~\cite{hapmap3} and 1000 genomes (1000G) projects~\cite{1000g}. 
\begin{figure}[htp]
\centering
\includegraphics[width=14cm]{\dir/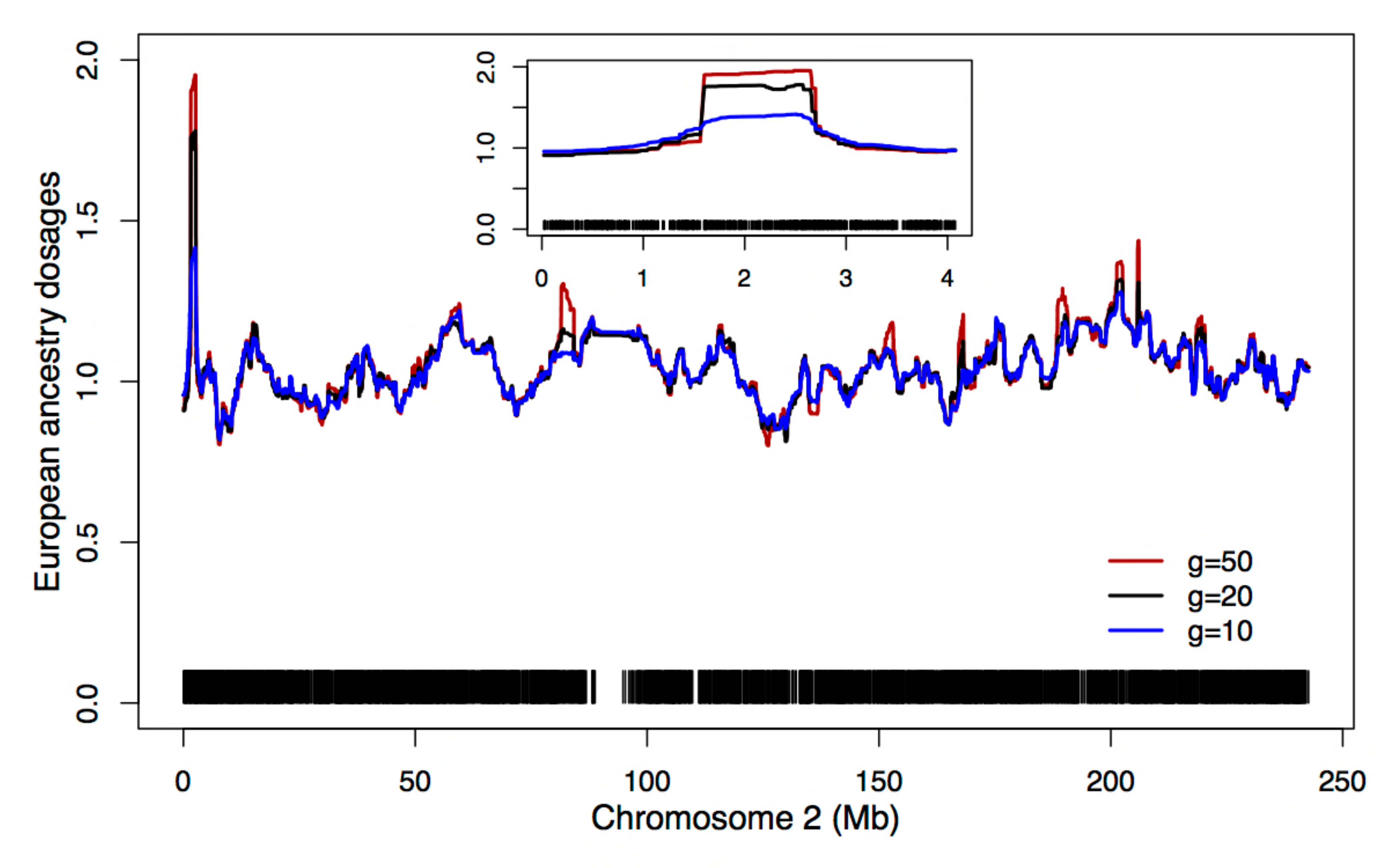}
\caption{Excessive local European ancestry in Mexicans. The y-axis is the average European ancestry allele dosages over $58$ Mexican samples. Black segments indicate SNPs. The inlet is a close-up of the peak. The peak region is in 2p25.3, ranging from 1.59Mb to 2.70Mb; the average European allele dosage is $1.95$.}
\label{fig:local-enrich}
\end{figure}

\medskip
{\bf HapMap3 Samples.}
We used $112$ individuals from CEU and $147$ individuals from YRI in HapMap3 and  $35$ individuals from Mayan and Pima in the Human Genetic Diversity Panel (HGDP)~\cite[][]{hgdp} as three source populations (denote as SP1) to infer the local ancestry of $58$ Mexican samples from HapMap3 (all genotypes are diploid). We fit the model with $S=3, K=15$ and $\g=10, 20$ or $50$ on Chromosome $2$. The mean ancestry proportions for CEU, YRI and Native Americans are $0.517, 0.048$ and $0.435$, respectively, in line with what has been reported by others~\cite{hua.11,multimix}. In examining local ancestral allele dosages, we found a $1.1$Mb region that contains excessive European ancestry (Figure~\ref{fig:local-enrich}). Note that because $1.1$ Mb is roughly equal to $1.1$cM, the result obtained using $\g=50$ has the strongest signal.  Interestingly,  this region harbors two genes, PXDN and MYT1L, both of which are associated with autism and schizophrenia~\cite{pxdn1,pxdn2}. Given that the prevalence of autism is lower in Hispanic children compare to other ethnic groups~\cite{autism1,autism2}, further studies are warranted to clarify whether and how this region contributes to the lower prevalence of autism in this population. 


Encouraged by this finding, we further analyzed all autosomes and identified four additional genetic regions (Supplementary Fig.~S3) that exhibit excessive European ancestry dosages (EAD). We chose EAD of $1.60$ as a cutoff because it represents roughly $6$ standard deviations away from the mean based on a conservative estimate assuming binomial sampling. Noticeably, 1) all five regions are short, ranging from $0.30$ to $1.26$Mb; 2) all are coding regions, many genes in these regions are immune related; and 3) these regions contain multiple hits from genome-wide association studies -- undoubtedly, our findings will help to interpret results and to design follow-up studies.  It is worthwhile to note that because these regions are short, only a method with high resolution can detect them. (This might explain why this interesting phenomenon has never been reported before.)

\begin{figure}[htp]
\centering
\includegraphics[width=7cm]{\dir/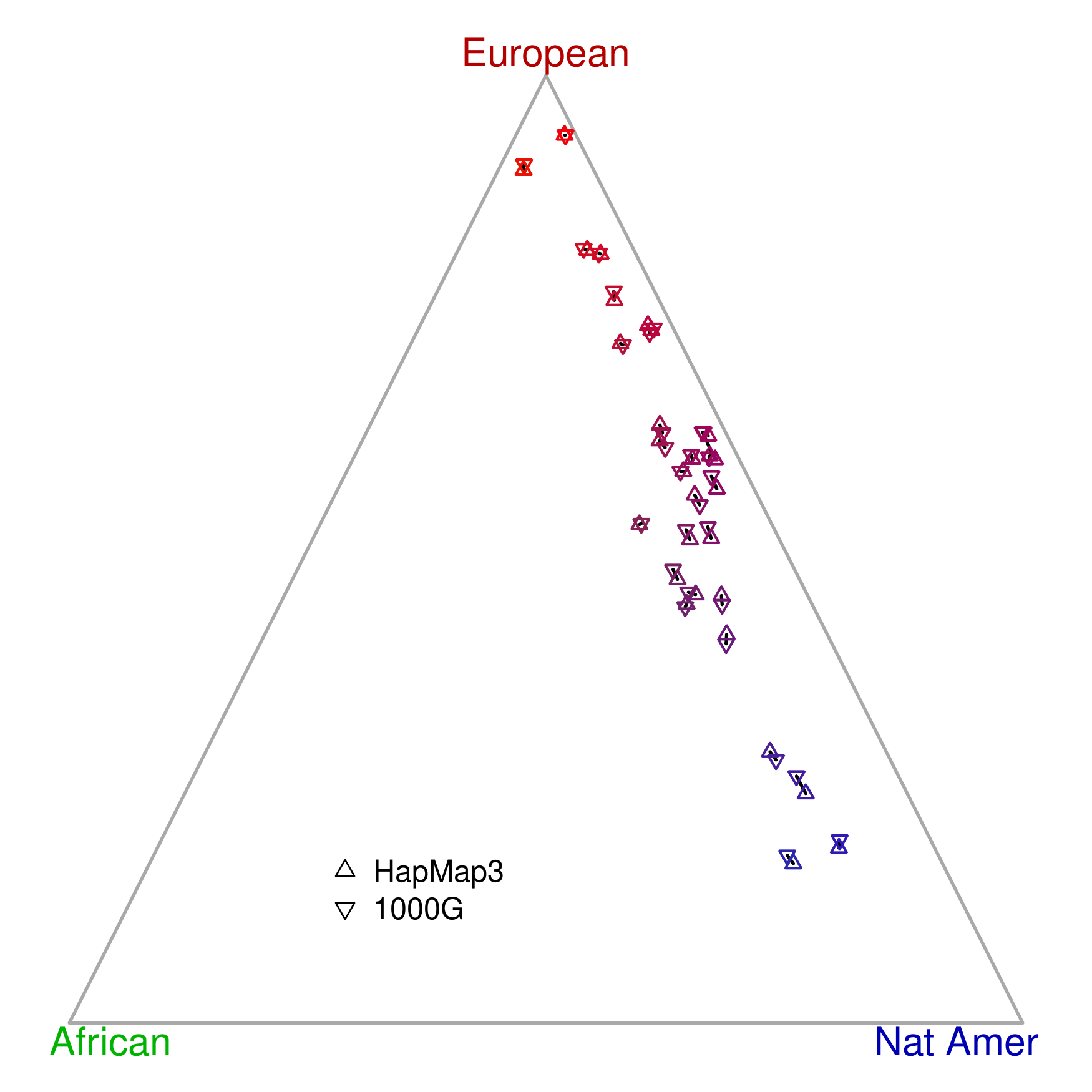}
\includegraphics[width=7cm]{\dir/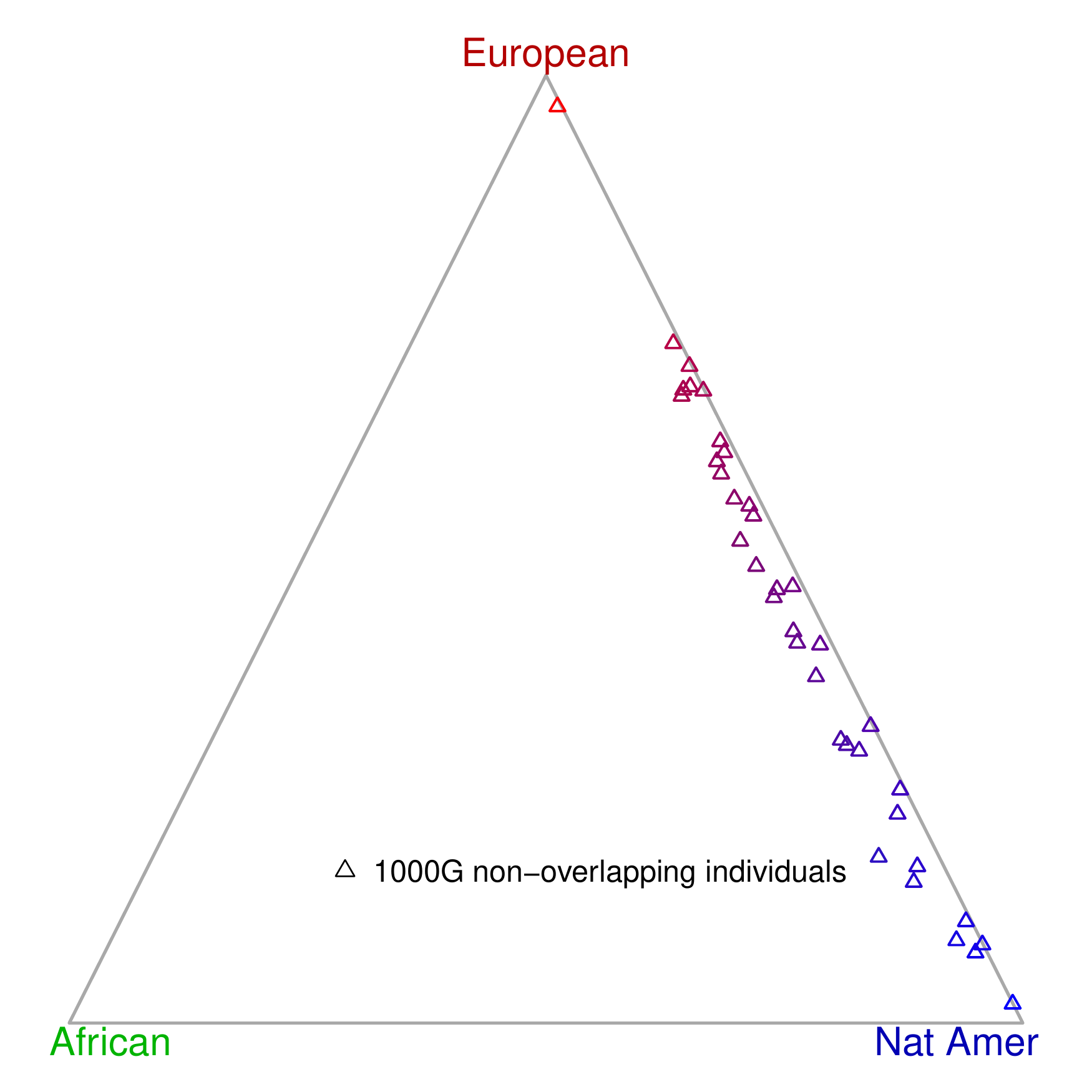}
\caption{Admixture proportions of 1000G Mexican samples (Chromosome 2). The left plot shows the concordance of $29$ overlapping individuals genotyped in HapMap3 and 1000G. Two points that belong to a same individual are connected by a short segment. The right plot shows the remaining $37$ Mexican samples in 1000G. Each individual has inferred admixture proportions, a triplet $(x,y,z)$ with $x+y+z=1$. An unique point can be determined when each component represents distance to an edge of an equilateral triangle. }
\label{fig:mex}
\end{figure}

\medskip
{\bf 1000G Samples.}
To double check our findings, we analyzed Mexican samples in the 1000G.  Using identity by state, we identified that $29$ of the total $66$ samples overlap with HapMap3 Mexican samples. For SNPs that are typed in both projects, there is a high genotype concordance for all $29$ samples (average Hamming distance $<0.002$). We inferred the local ancestries of these $66$ samples,  using haplotypes of CEU and YRI in 1000G and genotypes of Maya and Pima in HGDP as three source populations (denote as SP2).  We found that: 1) four among five regions were also discovered using these $66$ samples; the EAD of the region on chromosome $12$ dropped from $1.72$ to $1.49$, and EAD of a region on chromosome $18$ increased from previously $1.50$ to $1.60$ (see Supplementary Fig.~S3 for the region);  2) among $29$ overlapping individuals, the inferred admixture proportions have a high concordance  between two choices of source populations SP1 and SP2~(Figure~\ref{fig:mex}). Because we used unphased CEU and YRI in HapMap3 as source populations (SP1) for HapMap3 Mexican samples and used phased CEU and YRI  in 1000G  as source populations (SP2) for 1000G Mexican samples,  this high concordance suggests that the phasing of CEU and YRI in 1000G is reliable; and 3) the $37$ non-overlapping individuals in 1000G have an average smaller European ancestry proportion of $41.9\%$ compared to $56.6\%$ of those $29$ overlapping individuals (Figure~\ref{fig:mex}), and this difference is unlikely caused by random sampling (permutation test $p<0.004$).    

\begin{figure}[htp]
\centering
\includegraphics[width=15cm]{\dir/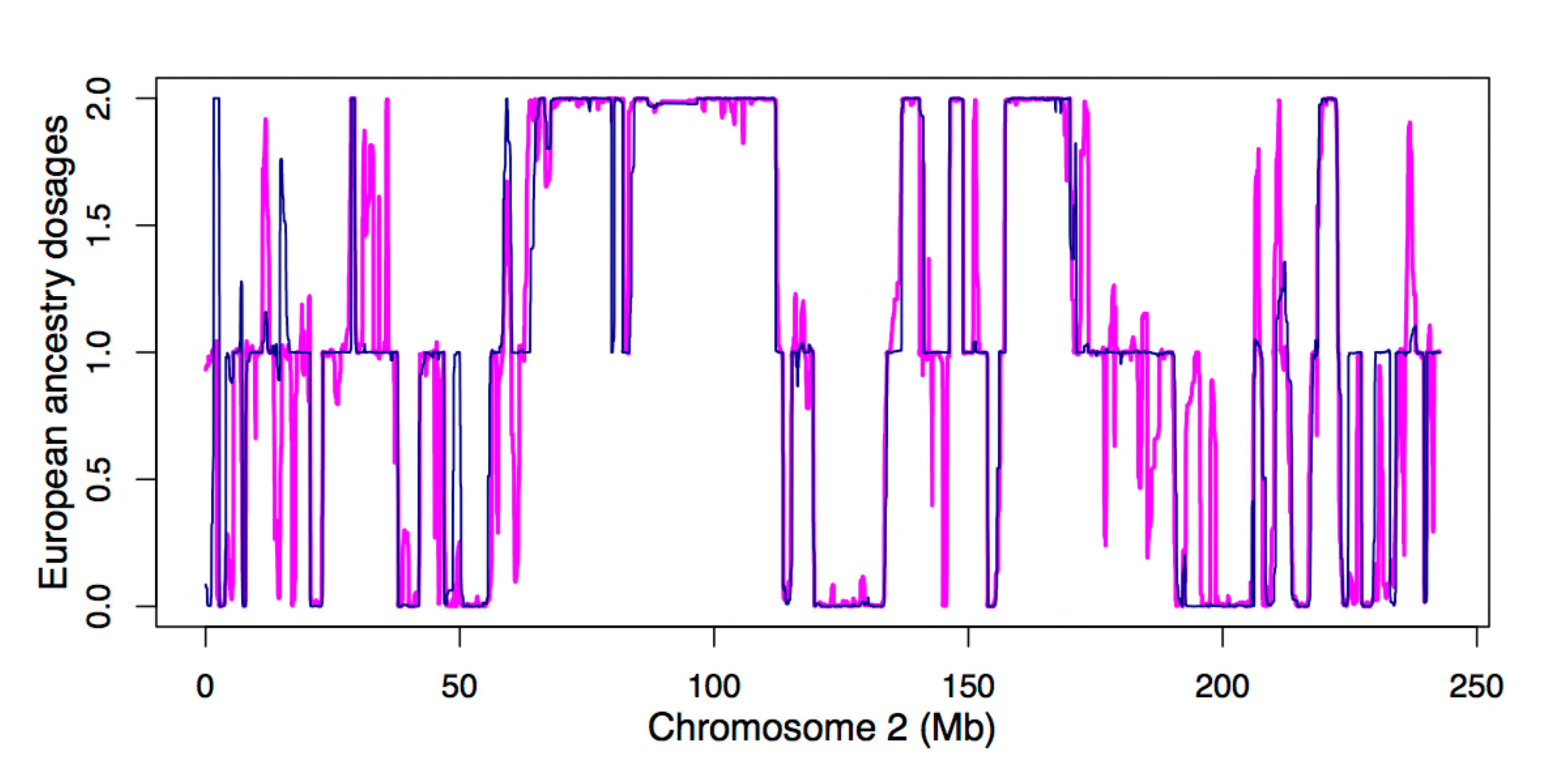}
\caption{Comparison between phased and unphased 1000G data. The plot shows the inferred European ancestry allele dosages (y-axis) of a typical Mexican individual. The x-axis denotes SNPs. The blue (pink) line denotes inferred values using unphased (phased) 1000G data. The excessive ancestry switches of pink line indicates imperfect phasing.}
\label{fig:unphase-phase}
\end{figure}

Since 1000G provides phased haplotypes for Mexicans, we therefore inferred the local ancestries of these haplotypes, using three source populations SP2. We found that the inferred local ancestries have excessive ancestry switches compared to those using unphased genotype data (Figure~\ref{fig:unphase-phase}).  These excessive switches are likely caused by imperfect phasing -- when using diploid genotypes our method integrates out phase uncertainties. Phasing admixed individuals is a hard problem.  Our results suggest, from an indirect angle,  that there is room for improvement in this area and we anticipate the two-layer model to make meaningful contributions.   

\section{Discussion} \label{sec:discussion}

We have presented a two-layer HMM to detect structure of local haplotypes, and demonstrated its utilities in local ancestry inference. Our method can directly work with diploid data and thus eliminates phase uncertainty that often plagues other methods.  The prevailing model for admixture is one pulse model, meaning haplotypes from two source populations mixed once some generations ago and continue to admix afterwards without influx of additional haplotypes from source populations. In reality, however, this assumption is overly simplified. Treating the mixing generation as a parameter, the two-layer model can average results over multiple choices of mixing generations. This makes our method applicable to the scenario of continuously mixing, which is perhaps a more realistic model for admixture. More importantly, our method has a high resolution -- owning to flexible choices of mixing generation, it is able to detect ancestry segments of $1$cM or smaller as demonstrated in the Mexicans data analysis.

Because structure of haplotypes is an ubiquitous phenomenon in genetic data, the two-layer model has many other potential applications. 
1) Using lower cluster dosages we can compute pairwise local haplotype sharing (LHS), defined as the probability of two haplotypes descent from a same lower cluster, which reflects genetic relatedness among haplotypes.  Preliminary studies suggest that LHS can be used to impute HLA alleles and detect genetic associations.  
2) As the two-layer model can infer the local ancestry with high accuracy,  it is reasonable to speculate that it will also be effective in genotype imputation and phasing for admixed individuals.  
3) Our method can directly estimate cluster-switch rates between adjacent markers, and this permits the inference of recombination rates and hotspots,  which will be particularly useful for admixed individuals. 
4) Aggregating is an effective method for detecting rare variants associations~\cite{li.leal.08}.  For admixed individuals, it would be helpful to aggregate rare variants of the same local ancestries.

Last, because a diploid individual has two sets of latent states (one for each haplotype), our EM algorithm is quadratic in both $S$ and $K$, and linear in numbers of individuals and markers.  This potentially limits the two-layer model's applicability. For local ancestry inference, when one can use phased data in source populations, the computation is fast because for a haploid individual, our EM algorithm is linear in $S$ and $K$.  Finding an appropriate linear approximation to fit our model for diploid individuals is a challenge and we are actively investigating this problem.  The recent progress concerning linear algorithms to fit the PAC model~\cite{shapeit} is extremely encouraging.  
Note that this quadratic computational challenge might disappear in the near future due to the recent development of methods such as phase-seq~\cite{phase-seq}, which produces genomic sequences completely phased across the entire chromosome.

\section{Acknowledgments}
The author would like to thank P. Scheet for helpful discussions regarding the $\theta$ update used in~\cite{fastphase} 
and A. Renwick for results of HAPMIX.  
M. Stephens and J. Belmont read and commented an early version of the paper.  

\bibliographystyle{unsrtnat}
\bibliography{2layer-ref,career-ref}

\newpage

\section*{Methods}

For ease of presentation, we assume that haploid individuals are being observed. Treating a diploid individual as having two random haploids subject to genotype constraint, our model applies directly to diploid individuals (Supplementary Note).  

\subsection*{The two-layer HMM}
We assume the numbers of clusters in upper and lower layers, $S$ and $K$, respectively, are fixed and the number of haplotypes is $N$ and number of markers is $M$.
For each individual $i$, let $X_m^\ii, Y_m^\ii$ be the latent state of the upper and lower clusters at locus $m$.    
Here $X_m^\ii$ and $Y_m^\ii$ take integer values to denote different clusters; each lower (upper) cluster associates with a parameter $\theta_{m\cdot}$ ($\eta_{m\cdot}$), representing ancestral allele dosages.  We may drop the superscript when referring to an arbitrary individual. 
\medskip

{\bf The main HMM.} 
The emission of an observed haplotype marker from a lower layer cluster is modeled as  
\begin{equation}
\begin{aligned}
p(h_m^\ii|X_m^\ii, Y_m^\ii, \xi) = p(h_m^\ii|Y_m^\ii, \xi)  & =  \left\{
\begin{array}{l l}
 \theta_{m\: Y_m^\ii} & \mbox{if $h_m=1$ } \\
1- \theta_{m \:Y_m^\ii} & \mbox{if $h_m=0$ } \\
 1 & \mbox{if $h_m$ is missing }  \\
\end{array}
\right .           	 
\end{aligned}. 
\end{equation}
The complete data likelihood has the form 
\begin{equation}
p(h^{(1)}, \dots, h^{(N)}, X^{(1)}, Y^{(1)}, \dots, X^{(N)},Y^{(N)} | \Theta)
= \prod_{i=1}^N \prod_{m=1}^M {p(h_m^\ii|Y_m^\ii, \xi) \; p(X_m^\ii, Y_m^\ii | \Theta)}.  
\label{eqn:likelihood:lower}
\end{equation}
The transition of the Markov chain on latent states is modeled as 
\begin{equation}
\begin{aligned}
P&(X_{m}=s, Y_{m}=k | X_{m-1} =s', Y_{m-1}=k')  \\ &=
j_{m}  \alpha_{s} \beta_{msk} +
(1-j_{m}) r_{m} \beta_{msk} I(s=s') +
(1-j_{m}) (1-r_{m}) I(s=s')I(k=k')          	 
\end{aligned}
\end{equation}
where $I(a=b)$ is an indicator function and  $j_\cdot$ and $r_\cdot$ are cluster-switch probabilities for the upper and lower layers respectively; and
\begin{equation}
\begin{aligned}
p(X_1=s, Y_1=k) &= p(Y_1=k| X_1=s) p(X_1=s) = \alpha_{s} \beta_{1sk}, 
\end{aligned}
\end{equation}
where $\alpha_\cdot$ is an individual specific $S$-vector to denote the admixture proportion, and $\beta_{m\cdot\cdot}$ is an $S\times K$ matrix shared by all individuals. 

We made three assumptions on the transition matrix of the hidden states. First, conditional on switch, the cluster to land at locus $m$ is independent of cluster at locus $m-1$. This assumption, used by previous models \cite{fastphase,li.stephens.03}, reduces the number of parameters and simplifies computation compare to specifying a whole transition matrix.
 Second, conditional on switch, the cluster to land is homogenous across loci and individual specific at the upper layer but heterogeneous across loci and shared among individuals at the lower layer. 
 The assumption on the upper layer makes $\alpha_\cdot$ naturally admixture proportions; the assumption on lower layer accommodates two facts: 1) LD patterns are heterogeneous across loci and 2) LD is a group property. 
Third, we assume that if the upper layer switches, then the lower layer must switch, and the lower layer might switch if the upper layer stays. This encourages upper layer-specific LD patterns.  

In the context of the local ancestry inference, the aforementioned transition explains how the two-layer model captures two scales of LD. A haplotype copies mosaically from (ancestral) haplotypes in one source population, and then may switch to another source population and copies mosaically from its haplotypes. The upper-layer switch probabilities $j_\cdot$ determine how frequently switches occur between different source populations and the lower layer switch probabilities $r_\cdot$ determine how frequently switches occur between ancestral haplotypes within each source population. Thus, the model can accommodate the two scales of LD observed in admixed individuals.      

In the main HMM, the upper layer latent state $X_m$ only contributes to transitions of latent states (through $\beta$),  not to emitting a marker or $\theta$ estimates (likelihood involves no $\eta$). This works well when $K$ is not too large, but it works less well for large values of $K$. To stabilize the estimates of $\theta$ for large values of $K$, we use an ancillary HMM to model $\eta$ emitting $\theta$. 

\medskip

{\bf The ancillary HMM.}   
We model emission of $\theta_{mk}$ from an upper-layer cluster $W_m^\kk$ as 
\begin{equation}
p(\theta_{mk} | W_m^\kk, \xi) = Beta(\theta_{mk}; F \eta_{m\:W_m^\kk}, F (1-\eta_{m\:W_m^\kk})), 
\end{equation}
where $Beta(x; a, b)$ denotes a Beta density with parameters $a,b$. This emission is adapted from the Balding-Nicols model~\cite{balding.nichols.95}. The original model is designed to model population divergence, and hence, $F$ is specified through Fst values between different populations.  In this situation, we use it as a ``random effect model" to stabilize $\theta$ estimates. For computational convenience, we set $F=1$ (Supplementary Note).  
We treat $\theta_{mk}$ as observed, and the complete data likelihood has the form
\begin{equation}
p(\theta_{\cdot 1}, \dots, \theta_{\cdot K}, W^{(1)}, \dots, W^{(K)} | \Theta)
= \prod_{k=1}^K \prod_{m=1}^M {p(\theta_{mk}|W_m^{(k)}, \xi) \; p(W_m^{(k)} | \Theta)}.  
\label{eqn:likelihood:upper}
\end{equation}
The transition of the latent states is modeled as 
\begin{equation}
\begin{aligned}
p(W_1^\kk=s) & = a_s^\kk \\
p(W_m^\kk = s | W_{m-1}^\kk = s') & = \rho_m a_s^\kk + (1-\rho_m) I(s=s'). 
 \end{aligned}
\end{equation}
We assume the jump probabilities $\rho_\cdot$ are unrelated to the jump probabilities of $j_\cdot$ and $r_\cdot$.  
 
\subsection*{Model fitting} 
 In the main HMM, the collection of parameters $\xi$ contains $\theta$ (an $M \times K$ matrix),  $\beta$ (an $M \times S \times K$ matrix) , $\alpha$ (an $N \times S$ matrix), and $j$ and $r$  (both $M$ vectors). In the ancillary HMM, the set of parameters contain $\eta$ (an $M \times S$ matrix), $a$ (a $K \times S$ matrix) and $\rho$ (an $M$ vector).   We briefly discuss how to estimate these parameters using Expectation Maximization (EM), focusing on the main HMM. For details, please refer to the Supplementary Note. 

For an arbitrary individual $i$, we write the forward probability $\phi(m, s, k) = p(h_{1:m}, X_m=s, Y_m=k|\xi)$ and backward probability $\psi(m, s, k) = p(h_{m+1:M} | X_m = s, Y_m = k|\xi)$. 
Both the forward and backward probabilities can be computed analytically. Then we obtain the posterior probability of latent states at each marker $p(X_m = s, Y_m = k| h, \xi) = \phi(m,s,k) \psi(m,s,k) / {p(h|\xi)}$ for each individual.  This allows us to compute quantities to update each parameter. 
Similarly, we can also update parameters in the ancillary HMM. 
Note that both $h$ and $\eta$ contribute to $\theta$ updates. 
Upon convergence of EM, we have $\xi^*$. 

\medskip
{\bf Constraint on cluster switches.}
Estimating switch probabilities $j$ and $r$ is trickier than others for two reasons. First, their updates are sticky, as a large $j_m$ (or $r_m$) estimate in a previous iteration often results in a large estimate in the current iteration. Because of this stickiness, the choice of initial values of $j$ and $r$ heavily affects the point at which they converge.   Second, $j$ and $r$ are not completely identifiable. If $\alpha_{s} \beta_{msk}$ concentrates on a particular cluster pair $(s,k)$, then a large probability in either $j_m$ or $r_m$ results in a similar likelihood.  We overcome these difficulties by putting constraints on $j$ and $r$ -- the coalescent intuition behind the two-layer model come to assist. 

Let $r_m = 1-\exp(-\lam_m^{(l)})$, where $\lam_m^{(l)} =  4N_ec_m\:\delta_l$ is the lower-layer cluster switch rate, where $N_e$ is the effective population size and $c_m$ is the genetic distance between markers $(m-1)$ and $m$.  We assume that $c_m$ is known. In practice, we approximate it using $1$megabase(Mb) per cM. 
Recalling that $T_k = 1/\binom{k}{2}$ is the mean coalescent time  for $k$ lineages, we have $\delta_l = T_{K+1} + \dots +T_N = \frac{1}{K}-\frac{1}{N}$.  Assume $N>> K$, and we have $\delta_l \approx1/K$.   
This leads to a natural choice for constraint on $\lam_m^{(l)}$ (and hence $r_\cdot$). For example, if $N_e=10,000$, $\sum c_m=5$cM and $K=10$, then we may apply constraint $\sum \lam_m^{(l)} = 500$. In practice, we directly estimate $r_m$ and compute $\lam_m^{(l)}$, and then we rescale $\lam_m^{(l)}$ to match the constraint and reestimate $r_m$. 

Let $j_m = \exp(-\lam_m^{(u)})$, where $\lam_m^{(u)} = 4N_ec_m\:\delta_u$. We are tempted to follow a similar coalescent argument to specify $\delta_u$.  However, unlike $\delta_l$, which is robust to recent demographic history because it pertains to an ``ideal" ancestral population, $\delta_u$ is  heavily influenced by demographic histories (for example, admixture generations) and the coalescent argument becomes ineffective.  To work around this issue, we constrain $j_\cdot$ through the \emph{admixture generation} $\g$.  In practice, we first estimate $j_m$ and compute $\lam_m^{(u)}$, and then we use $\sum \lam_m^{(u)} = \g \sum c_m$ to rescale $\lam_m^{(u)}$ to reestimate $j_m$.  
Define \emph{ancestry track length} $\l =M/\sum{\lam_m}$, then $\l$ and $\g$ follow a simple relationship $\g\; \l = 100$.

\medskip
{\bf Inference and computation.} 
We are interested in the \emph{upper cluster dosage} $P(X_m^\ii | h^\ii, \xi^*)$ for each individual $i$, which represents the local ancestry; its average represents the \emph{admixture proportion}. 
After trial and error, we arrived at the following model-fitting tricks that improve performances:
1) Because the dimension of $\xi$ is high and standard EM procedures tend to converge to a local mode instead of the global mode, it is useful to average inferences over multiple EM runs. 
2) It is helpful to initialize parameters with values that best preserves symmetry, e.g.,~$\theta_{mk} \approx 0.5$, $\alpha^\ii \approx 1/S$, and $\beta_{msk} \approx 1/K$ for all values of $m, s$ and $k$, respectively. 
The initial values can be simulated from symmetric Beta or Dirichlet distributions with large rates. 
3) The training data from source populations can be either phased or unphased. The difference is small when phasing is accurate and the computation with phased data is faster (linear vs. quadratic in terms of $S$ and $K$).  However, when phasing is less accurate, for example, pure statistical phasing without help of transmission, using unphased data is preferred. 
4) The common practice used in imputation~\cite{guan.stephens.08}, for which one first fits the model to the training data from source populations and then runs forward-backward algorithm once on the admixed individuals, tends to produce spurious ancestry switches in spikes.  Performing additional EM steps using both training data from source populations and admixed individuals (joint model fitting) reduces spurious ancestry switches. We recommend joint model fitting. 

\newpage

\setcounter{figure}{0}
\makeatletter 
\renewcommand{\thefigure}{N\@arabic\c@figure}

\section{Supplementary Note}

\subsection{Model Motivation}
Suppose we observe $N$ haplotypes $h^{(1)}, \dots, h^{(N)}$.  For a nonrecombined region, these $N$ haplotypes relate to one another through coalescent trees (Fig.~N1 for a typical $\tau$). 
We are interested in computing a joint likelihood $p(h^{(1)}, \dots, h^{(N)}|\Theta)$, where $\Theta$ is a collection of parameters to be specified later. Traditionally this is done by first sampling a set of $\tau$ and then conditional on each $\tau$ integrating out ancestral state $\theta$ using a ``peeling algorithm" \cite[c.f.][]{Fbook}.   
This computation involves sampling trees and this process can be extremely slow (and becomes much slower when considering recombinations). 
\begin{figure}[htp]
\centering
\includegraphics[width=8cm]{\dir/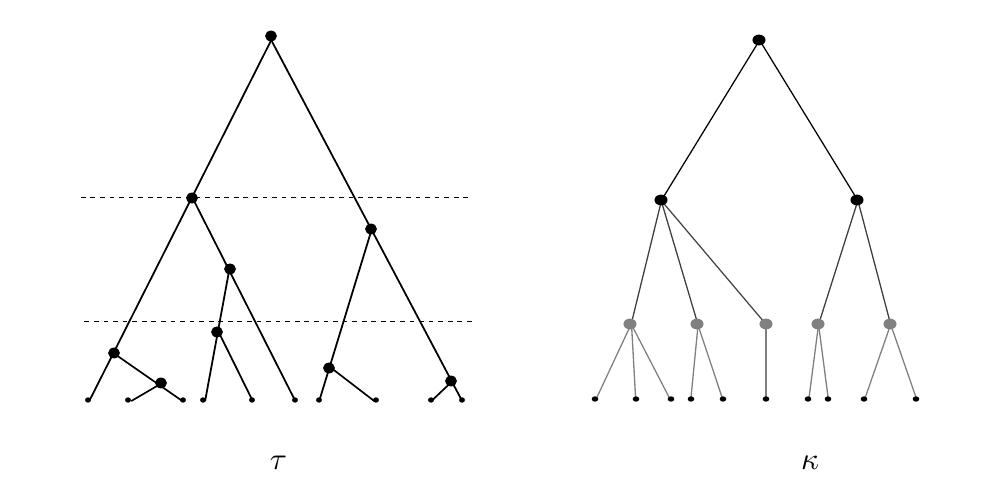}
\caption{Approximate coalescence. A coalescence tree $\tau$ on the left is cut at two different levels. The subtrees below each cut are approximated by star-trees to obtain $\kappa$ on the right.}
\label{fig:model}
\end{figure}
To make the joint likelihood computable for modern genetic data (e.g.,~genetic data from genome-wide association studies), we approximate the coalescent tree with a simpler structure -- two layers of clusters (not counting the root and leaf layers, Fig.~N1). 
We make two cuts on a tree and group nodes according to subtrees to form clusters.  The first cut is near the tip and individual haplotypes are grouped into lower clusters; the second cut is near the root and the lower clusters are grouped into upper clusters. 
Because a coalescent tree is a random tree, we allow each lower cluster to have positive probabilities of connecting to any upper clusters, and each individual haplotype to have positive probabilities of connecting to any lower clusters. Thus, $\kappa$ can represent not just one but many realizations of coalescent trees. We learn those probabilities from the data.  Cutting at different places will result in different numbers of clusters. In our HMM model, we prespecify the number of clusters in each layer and then cut the tree at places to match the number of clusters.  

Approximating $\tau$ with $\kappa$ allows us to condition on ancestral states integrate out $\kappa$, and then integrate out the ancestral states.  
Because $\kappa$ is more regular, it can be represented by latent states and recombination can be conveniently approximated by cluster switching within each layer.  These give us a two-layer HMM that greatly simplifies computation.

\subsection{Choice of parameters}
The companion software is easy to use -- users only need to specify three parameters: $S, K$ and $\g$.  
For local ancestry inference, $S$ usually is clear \emph{a priori}. 
We used $K=5S$ in our study, but the method is robust to a wide range of $K$ values. We demonstrate this through examples.  For a set of simulated two-way admixed individuals, we used $S=2$ and $K=5, 10$ or $20$ to fit the model. The results indicate that $K=10$ or $20$ outperforms $K=5$, especially when using correlation as a metric. However, the difference between $K=10$ and $K=20$ is small (Fig.~N2).  

As a rule, we recommend averaging results over multiple choices of $\g$. 
However, in general, $\g=10$ for African American samples, $\g=20$ for Latinos, and $\g=100$ for Uyghurs appear to be good choices. 
In our simulation studies, the local ancestry inference is robust to $\g$ up to a multiple of $2$.  However, $\g$ affects the smoothness of the local ancestry inference. 
This can be best demonstrated through examples. We simulated two-way admixed individuals with admixture generation $\g=100$ and fit the model using $\g=50, 100$ and $200$ respectively (Fig.~N3).  For all individuals, we found that small values of $\g$ produces smoother local ancestry estimates than large values of $\g$.  Nonetheless, for all three choice of $\g$, the main ancestry blocks are well inferred. Taking one individual as an example, the three deviations for three choices of $\g$ are $0.067, 0.062$, and $0.092$, respectively, with $\g=100$ performing the best and $\g=200$ performing much worse than the other two, presumably because the metric is more sensitive to smoothness. In fact, if measured by Pearson's correlation, we obtain $0.934, 0.947,$ and $0.932$ respectively. As a comparison, the deviation for HAPMIX is $0.067$ and the correlation is $0.939$. Although similar quantitatively to our method, HAPMIX does miss a major ancestry block in the middle (Fig.~N3).

\begin{figure}
\centering
\includegraphics[width=7cm]{\dir/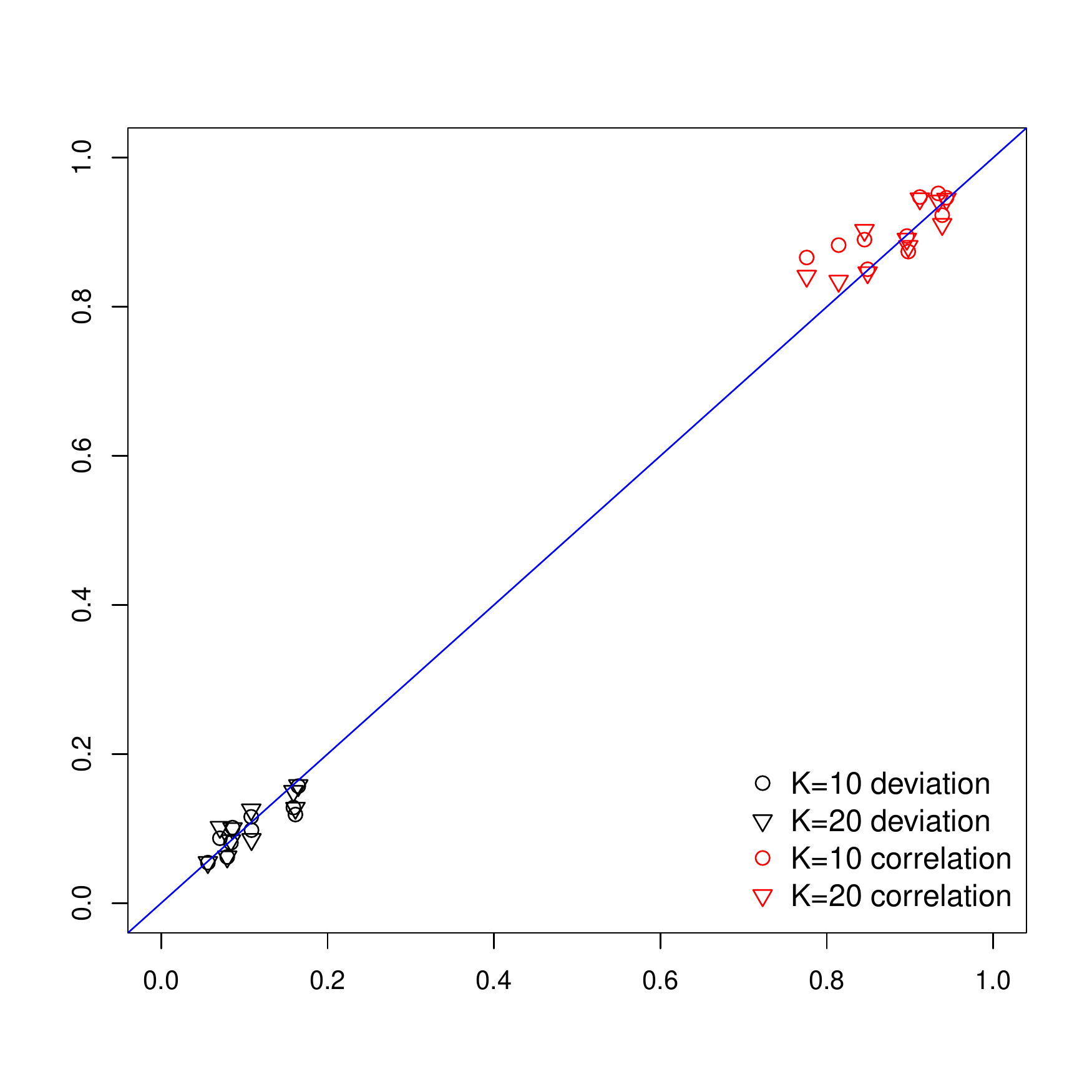}
\caption{Comparisons for different $K$. The plot shows comparisons between $K=5$ (on the x-axis) and $K=10$ or $20$ (on the y-axis). Black color denotes deviation and the red color denotes correlation between inferred and value and the truth along a chromosome. }
\label{fig:choicek}
\end{figure}

\begin{figure}
\centering
\includegraphics[width=7cm]{\dir/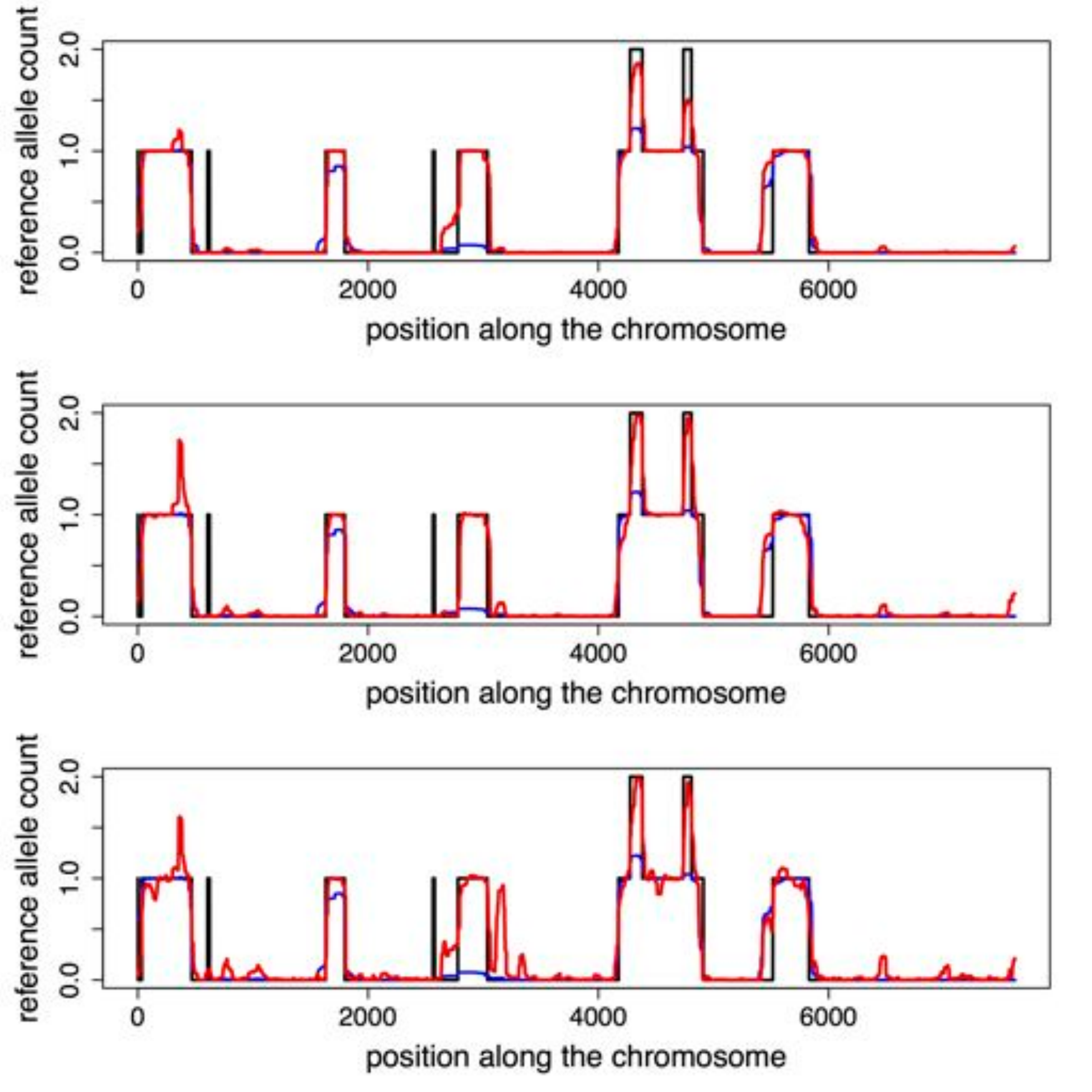}
\includegraphics[width=7cm]{\dir/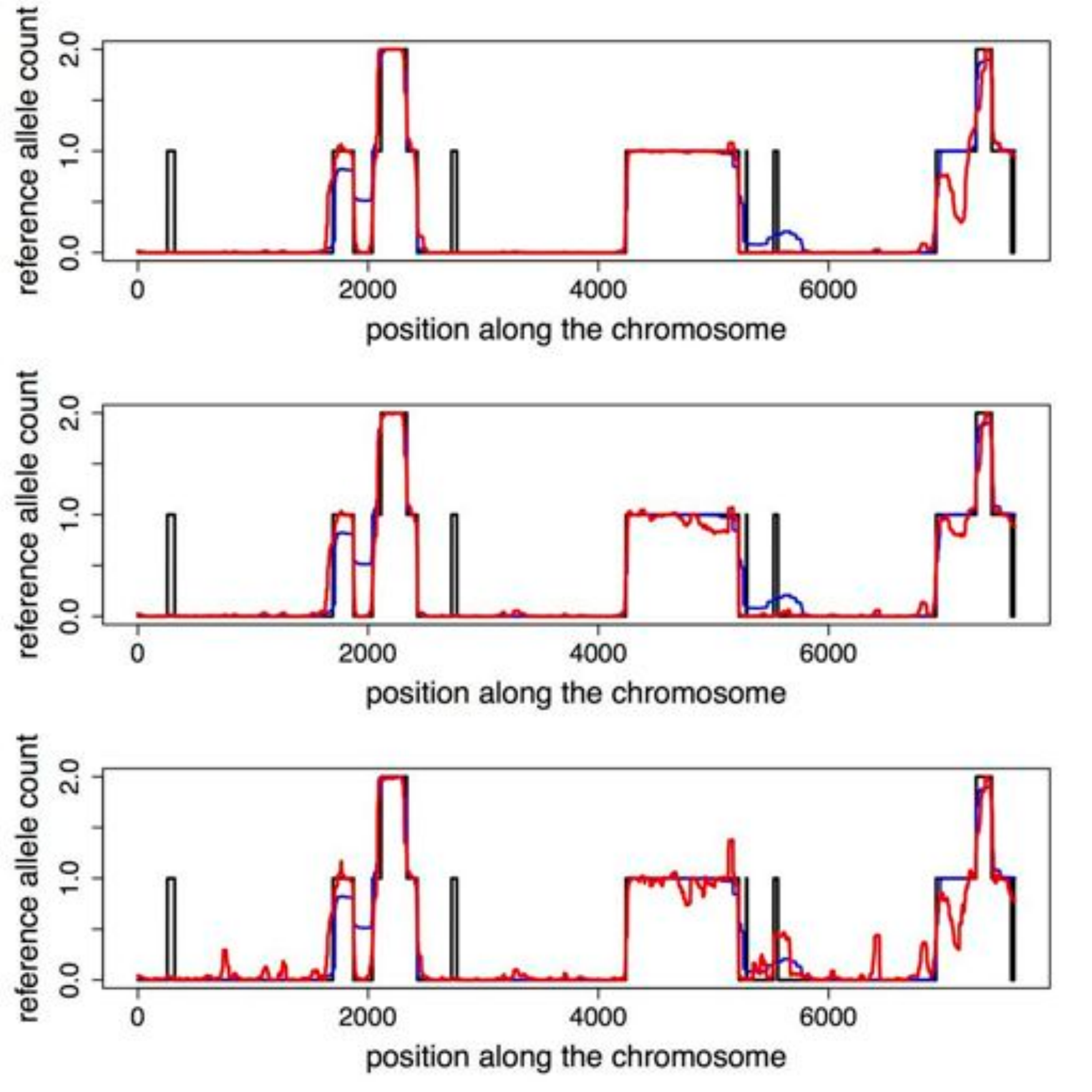}
\caption{Comparison for different $g$. The x-axis denotes genetic markers along a chrosome, the y-axis is inferred allele counts at each marker of an arbitrarily chosen ancestral population.The black lines are the simulated truth; the red lines are inferred values with different choice of mixing generations; the blue line are the results of the HapMix as a comparison. Each column denotes an individual, $\g=50,100,200$ from top to bottom panels. The individual on the left is explained in the main text. For the individual on the right, the deviation error are $0.053, 0.054, 0.077$, and correlations are $0.941, 0.946, 0.939$ respectively. As an comparison, HAPMIX has deviation $0.086$ and correlation $0.866$. }
\label{fig:choiceg}
\end{figure}

\subsection{Simulate admixed individuals} 
The procedure we used to simulate three-way admixed individuals is similar to what is used in HAPMIX~\cite{hapmix} for two-way admixture.  For a given admixture generation $\g$, we compute the average ancestral track length $\l = 100/\g$ and then compute $\lam = 1000 \l$ (a region of $1$Mb contains approximately $1000$ HapMap SNPs).  
We randomly choose three haplotypes, $h_c, h_y$ and $h_a$, from CEU, YRI, and CHB+JPT, respectively, and  
then copy from the three haplotypes to form a new admixed haplotype by repeating the following three steps: (1) let $s$ be the current position and generate a number $w$ according to an exponential distribution with mean $\lam$; (2) copy SNPs $(s,s+w]$ from $h_a$ with probability $\delta_1$, from  $h_y$ with probability $\delta_2$, and from $h_a$ with probability $\delta_3=1-\delta_1-\delta_2$; (3) increase s by $w$, finish if $s$ exceeds the total number of SNPs.  
Two admixed haplotypes are paired randomly to form a diploid individual. The markers are then thinned to match the Illumina 650K SNP chip. 
The two-way admixture can be simulated accordingly. We chose $(0.8, 0.2)$ as the target two-way admixture proportions and $(0.6, 0.2, 0.2)$ for three-way admixture proportions. Note that the simulated admixture proportions varies due to a finite number of SNPs.

\subsection{Expectation Maximization}

First outline the EM algorithm. Make an initial guess of parameters $\Theta^*$, 
and write down the complete data likelihood, denoting $Z_m^\ii =(X_m^\ii, Y_m^\ii)$, 
\begin{equation}
p(h^{(1)}, \dots, h^{(n)},Z^{(1)}, \dots, Z^{(n)}| \Theta^*) 
=  \prod_{i=1}^n \prod_{m=2}^M  {p(h_m^\ii|Z_m^\ii, \Theta^*)} p(Z_m^{(i)}| Z_{m-1}, \Theta^*)\; p(h_1^\ii|Z_1^\ii, \Theta^*) p(Z_1^\ii|\Theta^*).   
\end{equation}
Then estimate a new $\Theta$ by 
\begin{equation}
\begin{aligned}
\mbox{argmax}_{\Theta}\; E_{Z|h^{(1)}, \dots, h^{(n)}, \Theta^*}\left[\log p(h^{(1)}, \dots, h^{(n)},Z^{(1)}, \dots, Z^{(n)}| \Theta) \right]. 
\end{aligned}
\end{equation}
Set $\Theta^* = \Theta$ and iterate the procedure until $\Theta^*$ converges. 

To elaborate on the EM algorithm: conditioning on $\Theta^*$, the posterior distribution of $p(Z^\ii|h^\ii, \Theta^*)$ can be computed for each $i$.  To estimate $\Theta$, one can either sample many paths from $p(Z^\ii|h^\ii, \Theta^*)$ (hard EM), or one can integrate out $p(Z^\ii|h^\ii, \Theta^*)$ analytically (soft EM).  Intuitively, soft EM will perform better because it does not introduce sampling variation. However, with hard EM only forward probabilities need to be computed in order to sample from $p(Z^\ii|h^\ii, \Theta^*)$. More importantly, computational tricks may be applied on the sampled paths to avoid possible traps of local optimum. 
In the paper we use soft EM for model fitting and report possible computational improvement elsewhere. 
 
A diploid individual has two sets of latent states at each marker $Z_m^1=(X_m^1,Y_m^1), Z_m^2=(X_m^2,Y_m^2)$ to indicate the upper and lower layer cluster membership (we drop the super script for individual and this should cause no confusion). The conditional likelihood for $i$-th individual is $p(g^\ii|Z^1, Z^2, \Theta) = \prod_{m=1}^M{p(g_m^\ii|Y_m^1, Y_m^2, \Theta)}$ with  ``emission"
\begin{equation}
p(g_m^\ii|Y_m^1=j, Y_m^2=k, \Theta) =  \left\{
\begin{array}{l l}
t_{j} t_{k} & \mbox{if  $g_m^\ii = 2$ } \\
t_{j} (1-t_{k}) + (1-t_{j}) t_{k}  & \mbox{if  $g_m^\ii = 1$ } \\
(1-t_{j}) (1- t_{k}) & \mbox{if $g_m^\ii = 0$} \\
1 & \mbox{if $g_m^\ii$ is missing.}
\end{array}
\right.
\end{equation}
where 
\begin{equation}
\begin{aligned}
t_{j} &=\theta_{mj} (1-\mu) + (1-\theta_{mj}) \mu \\
t_{k} &=\theta_{mk} (1-\mu) + (1-\theta_{mk}) \mu \\
\end{aligned}
\end{equation}
and $\mu = 4N \nu$ is the scaled mutation rate. In the implementation we used $\mu = 0.001$.  Note the one-to-one correspondence between $t_{\cdot}$ and $\theta_{m\cdot}$, and that we implicitly assumed Hardy-Weinberg equilibrium in the emission.

\subsubsection{Forward and backward recursion} 
In what follows, every probability statement is conditional on $\Theta^*$. We assume there are $M$ biallelic SNPs. The forward recursion. 
\begin{equation}
\begin{aligned}
&\phi(m+1, s_1,k_1, s_2, k_2) =  p\left(g_{1:m+1}^\ii,  Z_{m+1}^1=(s_1, k_1), Z_{m+1}^2=(s_2,k_2) \big| \Theta^* \right) \\
&=p(g_{m+1}^\ii|Z_{m+1})  \sum_{s'_\cdot,k'_\cdot}\phi(m, s'_1,k'_1, s'_2,k'_2) p\left(Z_{m+1}^1| Z_m^1=(s'_1,k'_1)\right) p\left(Z_{m+1}^2| Z_m^2=(s'_2,k'_2)\right) \\
&=p(g_{m+1}^\ii|Z_{m+1})\; \left( j_{m+1}^2 \; p_{11} + j_{m+1} (1-j_{m+1}) (p_{10} + p_{01} ) + (1-j_{m+1})^2 \; p_{00}\right),
\end{aligned}
\end{equation}
where $\phi(1,s_1,k_1,s_2,k_2) = \alpha_{s_1}^\ii, \beta_{1,s_1,k_1} \alpha_{s_2}^\ii, \beta_{1,s_2,k_2}\;p(g_1^\ii|s_1,k_1,s_2,k_2)$ and 

\begin{equation}
\begin{aligned}
p_{00} &= (1-r_{m+1})^2  \phi(m, s_1,k_1,s_2,k_2)+
r_{m+1}^2  \sum_{k'_1,k'_2} \phi(m, s_1,k'_1,s_2,k'_2) \beta_{m+1,s_1,k_1} \beta_{m+1,s_2, k_2}  + \\
&r_{m+1} (1-r_{m+1}) \left( \sum_{k'_1} \phi(m, s_1,k'_1,s_2,k_2)  \beta_{m+1,s_1,k_1} +  \sum_{k'_2}\phi(m, s_1,k_1,s_2,k'_2) \beta_{m+1,s_2,k_2}   \right) 
\end{aligned}
\end{equation}
\begin{equation}
p_{10} = \alpha_{s_1}^\ii \beta_{m+1,s_1,k_1} \Big( r_{m+1}  \sum_{s'_1, k'_1, k'_2} \phi(m, s'_1,k'_1,s_2,k'_2) \beta_{m+1,s_2,k_2} + 
  (1-r_{m+1})  \sum_{s'_1, k'_1}  \phi(m, s'_1,k'_1,s_2,k_2)  \Big) 
\end{equation}
\begin{equation}
p_{01} = \alpha_{s_2}^\ii \beta_{m+1,s_2,k_2}  \Big( r_{m+1}   \sum_{s'_2, k'_1, k'_2}\phi(m, s_1,k'_1,s'_2,k'_2) \beta_{m+1,s_1,k_1}  + 
  (1-r_{m+1})  \sum_{s'_2, k'_2} \phi(m, s_1,k_1,s'_2,k'_2)  \Big) 
\end{equation}
\begin{equation}
\begin{aligned}
p_{11} &= \alpha_{s_1}^\ii \beta_{m+1,s_1,k_1}   \alpha_{s_2}^\ii \beta_{m+1,s_2,k_2}  \sum_{s'_1,k'_1,s'_2,k'_2} \phi(m, s'_1,k'_1,s'_2,k'_2). 
\end{aligned}
\end{equation}

All summation with dummy variables $s'_\cdot,t'_\cdot$ only needs to be done once. This is the benefit of the parameterization for Markov transition described in the paper. The overall complexity of the forward and backward recursion is $O(MS^2K^2)$ for diploid individuals and $O(MSK)$ for haploid individuals. 

The backward recursion.  Note $ p(g_{m:M}^\ii|s_1, k_1, s_2,k_2) = \psi(m, s_1,k_1,s_2,k_2) p(g_m^\ii|s_1,k_1,s_2,k_2)$. 
\begin{equation}
\begin{aligned}
&\psi(m-1, s_1,k_1, s_2, k_2) =  p\left(g_{m:M}^\ii |  Z_{m-1}^1=(s_1, k_1), Z_{m-1}^2=(s_2,k_2)\big|\Theta^* \right) \\
&=\sum_{s'_1,k'_1,s'_2,k'_2}\psi(m, s'_1,k'_1, s'_2,k'_2) p(g_{m}^\ii|s'_1, k'_1,s'_2,k'_2)  p\left(Z_{m}^1=(s'_1,k'_1)| Z_{m-1}^1\right) p\left(Z_{m}^2=(s'_2,k'_2)| Z_{m-1}^2\right) \\
&=\left( j_m^2\; q_{11} + j_m (1-j_{m}) (q_{10} + q_{01} ) + (1-j_{m})^2  q_{00}\right),   
\end{aligned}
\end{equation}
where $\psi(M, s_1, k_1, s_2, k_2) = 1$ and 

\begin{equation}
\begin{aligned}
q_{00} &= 
 r_m^2  \sum_{k'_1,k'_2} \beta_{m,s_1,k'_1} \beta_{m,s_2, k'_2} p(g_{m:M}^\ii|s_1,k'_1,s_2,k'_2) + (1-r_m)^2  p(g_{m:M}^\ii| m, s_1,k_1,s_2,k_2) \\ 
 +& r_m (1-r_m) \times
\Big[\sum_{k'_1} \beta_{m,s_1,k'_1} p(g_{m:M}^\ii|s_1,k'_1,s_2,k_2) + \sum_{k'_2} \beta_{m,s_2,k'_2} p(g_{m:M}^\ii|s_1,k_1,s_2,k'_2)  \Big]  
\end{aligned}
\end{equation}
\begin{equation}
q_{10} =  r_m   \sum_{s'_1, k'_1, k'_2}  \alpha_{s'_1}^\ii \beta_{m,s'_1,k'_1}  \beta_{m,s_2,k'_2} p(g_{m:M}^\ii | s'_1,k'_1,s_2,k'_2) + 
 (1-r_m)  \sum_{s'_1, k'_1}  \alpha_{s'_1}^\ii \beta_{m,s'_1,k'_1} p(g_{m:M}^\ii | s'_1,k'_1,s_2,k_2)  
\end{equation}
\begin{equation}
q_{01} =  r_m   \sum_{s'_2, k'_1, k'_2}  \alpha_{s'_2}^\ii \beta_{m,s'_2,k'_2}  \beta_{m,s_1,k'_1} p(g_{m:M}^\ii | s_1,k'_1,s'_2,k'_2) + 
 (1-r_m)  \sum_{s'_2, k'_2} \alpha_{s'_2}^\ii \beta_{m,s'_2,k'_2}  p(g_{m:M}^\ii| s_1,k_1,s'_2,k'_2) 
\end{equation}
\begin{equation}
\begin{aligned}
q_{11} &= \sum_{s'_1,k'_1,s'_2,k'_2} \alpha_{s'_1}^\ii \beta_{m,s'_1,k'_1}   \alpha_{s'_2}^\ii \beta_{m,s'_2,k'_2}  p(g_{m:M}^\ii | s'_1,k'_1,s'_2,k'_2). 
\end{aligned}
\end{equation}

The posterior of latent states at each locus for each individual can be computed via
\begin{equation}
p\left(Z_m^1=(s_1,k_1), Z_m^2=(s_2,k_2) |g^\ii, \Theta^*\right) \propto \phi(m, s_1, k_1, s_2, k_2)  \psi(m, s_1, k_1, s_2, k_2)
\end{equation}
and renormalize to have $\sum_{s_1,k_1,s_2,k_2}p\left(Z_m^1=(s_1,k_1), Z_m^2=(s_2,k_2) |g^\ii, \Theta^*\right) = 1$. 

\subsubsection{Update $\theta$}
To update parameters in each EM steps, we solve for each component $x$ of $\Theta$,
\begin{equation}
\frac{d}{dx} E_{Z|h, g, \Theta^*}\left[\log p(h, g, Z^{(1)}, \dots, Z^{(n)}| \Theta) \right]  = 0. 
\end{equation}
Assume we have both diploid $g$ and haploid $h$ individuals in our data. 
For diploid individuals, at locus $m$, write $p_{ijk} = \sum_{s_1,s_2}p\left(Z_m^1=(s_1, j), Z_m^2=(s_2,k) |g^\ii, \Theta^*\right)$. 
Let $S_k = \{i: g_m^\ii = k\}$ for $k=0,1,2$. Similarly, for haploid individuals, at locus $m$, write $q_{ij} =\sum_{s}p\left(Z_m=(s, j) |h_m^\ii, \Theta^*\right)$. 
Let $T_k = \{i: h_m^\ii = k\}$ for $k=0,1$.  Let 
\begin{equation}
\begin{aligned}
a_{0j\odot} = \sum_{i \in S_0, k\neq j}{p_{ijk}},\hspace{.1in}
&a_{0jj} = \sum_{i \in S_0}{p_{ijj}},  \\
a_{2j\odot} = \sum_{i \in S_2, k\neq j} {p_{ijk}}, \hspace{.1in}
&a_{2jj} = \sum_{i \in S_2} {p_{ijj}}, \\ 
a_{1jk} = \sum_{i \in S_1}{p_{ijk}},  \hspace{.1in}
&a_{1jj} = \sum_{i \in S_1}{p_{ijj}}, \\
b_{0j} = \sum_{i \in T_0}{q_{ij}},   \hspace{.1in}
& b_{1j} = \sum_{i \in T_1} {q_{ij}}.  
\end{aligned}
\end{equation}
Take derivative with respect to $\theta_{mj}$ and sum over $k$ for diploid individual to get
\begin{equation}\label{eqn:Fj}
F_j(t_{\cdot}) = \frac{-1 }{1- t_j} (a_{0j\odot}+2a_{0jj}+a_{1jj}+b_{0j}) + \frac{1}{t_j} (a_{2j\odot} +2a_{2jj}+ a_{1jj}+b_{1j}) + 
\sum_{k \neq j} \frac{1-2 t_k}{t_j+t_k - 2 t_j t_k} a_{1jk} = 0
\end{equation}
for each $j = 1 \dots, K$ (recall $K$ is the number of lower-layer clusters).  We have $K$ equations with $K$ unknowns and we can solve numerically for $t_j$ and hence $\theta_{mj}$. 
To do so, we need the Jacobian $J(t_{\cdot})=(d_{jk})$ where 
\begin{equation}
d_{jk} = \frac{d F_j}{d t_k} = \frac{-1}{(t_j+t_k - 2 t_j t_k)^2} a_{1jk} \mbox{ \hspace{.2in}  for $k \neq j$}, 
\end{equation}
and 
\begin{equation}
d_{jj} = \frac{-1 }{(1- t_j)^2} (a_{0j\odot}+2a_{0jj}+a_{1jj}+b_{0j}) + \frac{-1}{t_j^2} (a_{2j\odot} +2a_{2jj}+ a_{1jj}+b_{1j}) + 
\sum_{k \neq j} \frac{(1-2 t_k)^2}{(t_j+t_k - 2 t_j t_k)^2} a_{1jk}. 
\end{equation}
We can solve $J(t^{(n)})(t^{(n+1)} -t^{(n)}) = -F(t^{(n)})$ for the unknown $t^{(n+1)}-t^{(n)}$. 
 
Compare to the update used in~\cite{fastphase}, this update for $\theta$ does not directly involve its previous value. Perhaps unwilling to solve a linear system repetitively, \citet{fastphase} used approximation to the last terms of Equation \eqref{eqn:Fj}. 
\begin{equation}
 \sum_{k \neq j} \frac{1-2 t_k}{t_j+t_k - 2 t_j t_k} a_{1jk} =  \sum_{k \neq j} \left(\frac{1}{t_j} a'_{1jk} - \frac{1}{1-t_j} a''_{1jk}\right),
 \end{equation}
  where 
    \begin{equation}
\begin{aligned}
  a'_{1jk} = \frac{t_j ( 1- t_k)}{t_j+t_k-2t_jt_k} a_{1jk}, \hspace{.1in} a''_{1jk} = \frac{t_k ( 1- t_j)}{t_j+t_k-2t_jt_k} a_{1jk}, 
  \end{aligned}
\end{equation}
  which can be computed by approximating $t_j$ and $t_k$ with values in the previous iteration. Denote 
  \begin{equation}
\begin{aligned}
  a'_{1j\odot} = \sum_{k\neq j}{a'_{1jk}}, \hspace{.1in} a''_{1j\odot} = \sum_{k\neq j}{a''_{1jk}},  
\end{aligned}
\end{equation}
and we have, 
\begin{equation}
F_j(t_{\odot}) = \frac{-1 }{1- t_j} (a_{0j\odot}+2a_{0jj}+a_{1jj}+b_{0j} + a''_{1j\odot}) + \frac{1}{t_j} (a_{2j\odot}+2a_{2jj} + a_{1jj}+b_{1j} + a'_{1j\odot})  = 0
\end{equation}
and solve to get 
\begin{equation}\label{eqn:tj}
t_j=  \frac{(a_{2j\odot} + 2a_{2jj}+ a'_{1j\odot} +a_{1jj} + b_{1j}) } { (a_{0j\odot}+2a_{0jj}+ a_{1j\odot}+2a_{1jj} + a_{2j\odot} +2a_{2jj} +b_{0j}+b_{1j})}. 
\end{equation}
With \eqref{eqn:tj} as a starting point only a few iterations are needed to estimate $\theta$ using numerical method described earlier. Note however, solving the linear system has complexity $O(K^3)$, which makes the complexity of model fitting to be $O(\max{(MS^2K^2, MK^3)})$.  

\subsubsection{Update Markov transition parameters}
To estimate Markov transition parameters, following~\cite{fastphase},  we introduce latent state transitions (jumps) $J_{im}$ and $R_{im}$ occured between locus $m-1$ and $m$ at upper and lower layers for individual $i$. 
Denote $J_{ims}$ the number of upper layer jumps to $X_m = s$, and $R_{imsk}$ the number of lower layer jumps to $X_m^\ii=s$ and $Y_m^\ii=k$.   
Recognize that $J_{ims}$ and $R_{imsk}$ are sufficient for $\alpha, \beta, j$ and $r$, we have 
\begin{equation}
\begin{aligned}
\alpha_s^\ii &= \frac{\sum_{m=2}^M{E[J_{ims}|g^\ii,\Theta^*]}} {\sum_{m=2}^{M}\sum_{s}{E[J_{ims}|g^\ii,\Theta^*]}} \\
\beta_{msk} &= \frac{\sum_i{E[R_{imsk}|g^\ii,\Theta^*]}}{\sum_{i,k}{E[R_{imsk}|g^\ii,\Theta^*]}} \\
j_m &= \frac{\sum_{i,s}E[J_{ims}|g^\ii, \Theta^*]}{\mbox{Number of haploids}} \\
r_m &= \frac{\sum_{i,s,k}E[R_{imsk}|g^\ii, \Theta^*]}{\mbox{Number of haploids} \times S},  
\end{aligned}
\end{equation}
where recall that $S$ is the number of upper-layer clusters. 

In what follows, when a state in forward or backward probabilities was substitute by a dot, it means that component was summed over.  
Note that $p(g^\ii|\Theta^*) = \phi(M,\cdot,\cdot,\cdot,\cdot)$ and 
$$p(g_{m:M}^\ii|s,k_1, s_2, k_2, \Theta^*) = p(g_m^\ii|s_1,k_1,s_2,k_2,\Theta^*) \psi(m,s_1,k_1,s_2,k_2).$$  

First  
$E[J_{ims}|g^\ii, \Theta^*] = 2 p(J_{ism} =2|g^\ii, \Theta^*) + p(J_{ism} =1|g^\ii, \Theta^*) $
 with
  \begin{equation}
2 p(J_{ism} =2|g^\ii, \Theta^*) =  
 2  (\alpha_s^\ii j_m)^2  / p(g^\ii | \Theta^*) \times \phi(m-1, \cdot, \cdot, \cdot, \cdot) \sum_{k_1,k_2}\beta_{msk_1} \beta_{msk_2}  p(g_{m:M}^\ii|s,k_1, s, k_2, \Theta^*),
 \end{equation}
and
  \begin{equation}
 \begin{aligned}
p(J_{ism} =1&|g^\ii, \Theta^*) =  j_m (1-j_m)\alpha_{s}^\ii / p(g^\ii | \Theta^*) \times \\
\Big[& (1-r_m) \sum_{s_2,k_2}\phi(m-1, \cdot, \cdot,s_2, k_2) \sum_{k_1} \beta_{msk_1} p(g_{m:M}^\ii|s,k_1, s_2, k_2, \Theta^*) +\\
&r_m \sum_{s_2}\phi(m-1, \cdot, \cdot,s_2, \cdot) \sum_{k_1,k_2} \beta_{ms_2k_2}\beta_{msk_1} p(g_{m:M}^\ii|s,k_1, s_2, k_2, \Theta^*) + \\
& (1-r_m) \sum_{s_1,k_1}\phi(m-1,s_1, k_1,\cdot, \cdot) \sum_{k_2} \beta_{msk_2} p(g_{m:M}^\ii|s_1,k_1, s, k_2,\Theta^*)  +\\
&r_m \sum_{s_1}\phi(m-1, s_1, \cdot, \cdot, \cdot) \sum_{k_1,k_2} \beta_{ms_1k_1}\beta_{msk_2} p(g_{m:M}^\ii|s_1,k_1, s, k_2, \Theta^*) \Big] .
\end{aligned}
 \end{equation}

Second, 

  \begin{equation}
 \begin{aligned}
 E[R_{imsk}|g^\ii, \Theta^*] = 2 p(R_{imsk} =2,J_{ims}=0 | g^\ii, \Theta^*)& + p(R_{imsk} =1, J_{ims}=0 | g^\ii, \Theta^*) \\
 & + p(R_{imsk} =1, J_{ims}=1| g^\ii, \Theta^*), 
\end{aligned}
 \end{equation}
 with each component being
   \begin{equation}
 \begin{aligned}
 2 p(R_{imsk} =2, J_{ims}=0| g^\ii, & \Theta^*) =  2(1-j_m)^2 r_m^2\beta_{msk}^2  /p(g^\ii|\Theta^*) \times \\
& \phi(m-1,s,\cdot,s,\cdot) p(g_{m:M}^\ii|s,k,s,k,\Theta^*),
   \end{aligned}
 \end{equation}
   \begin{equation}
 \begin{aligned}
 p(R_{imsk} =1, J_{ims}=0| g^\ii,& \Theta^*) = (1-j_m)^2 r_m(1-r_m) \beta_{msk}  /p(g^\ii|\Theta^*) \times \\
\Big[ &\sum_{s_2,k_2} \phi(m-1,s,\cdot,s_2,k_2) p(g_{m:M}^\ii|s,k,s_2,k_2,\Theta^*) \\
 &+ \sum_{s_1,k_1} \phi(m-1,s_1,k_1,s,\cdot) p(g_{m:M}^\ii|s_1,k_1,s,k,\Theta^*)\Big], 
\end{aligned}
 \end{equation}
 \begin{equation}
 \begin{aligned}
 p(R_{imsk} =1, J_{ims}=1| g^\ii,& \Theta^*) = j_m(1-j_m) r_m \beta_{msk}  /p(g^\ii|\Theta^*) \times \\
\Big[&\phi(m-1,s,\cdot,\cdot,\cdot) \sum_{s_2,k_2} \alpha_{s_2}^\ii \beta_{ms_2k_2}p(g_{m:M}^\ii|s,k,s_2,k_2,\Theta^*)  \\
&+ \phi(m-1,\cdot,\cdot,s, \cdot) \sum_{s_1,k_1} \alpha_{s_1}^\ii \beta_{ms_1k_1}p(g_{m:M}^\ii|s_1,k_1,s,k,\Theta^*) \Big].
\end{aligned}
 \end{equation}

Finally, special treatment is needed at marker $m=1$.  For each $s, k$ set
 \begin{equation}
 E[R_{i1sk}|g^\ii, \Theta^*] = \alpha_{s} \beta_{1sk} p(g_{1:M}^\ii|s,k,\cdot,\cdot,\Theta^*)
 \end{equation}
 and renormalize such that $\sum_{s,k}E[R_{i1sk}|g^\ii, \Theta^*] = d$, where $d = 2, 1$ for diploid and haploid individuals respectively. Set $E[J_{i1s}|g^\ii, \Theta^*] = \sum_{k}E[R_{i1sk}|g^\ii, \Theta^*]$.

\subsubsection{Ancillary HMM}
The expected complete data log-likelihood is given below. 
\begin{equation}
\begin{aligned}
E_{W|h,g,\xi^*}\left[~\sum_{j=1}^K \sum_{m=1}^M {\log p(\theta_{mj}, W_m^{(j)} | \eta_{ms}, \xi^*)}~ \right] 
  = \sum_{m=1}^M \sum_{j=1}^K   {\log p(\theta_{mj} | \eta_{ms}, \xi^*) p_{mjs}} 
\end{aligned}
\end{equation}
where $p_{mjs}$ is the $s$-th upper cluster dosage of of $j$-th haplotype at marker $m$. From Balding-Nichols model~\cite{balding.nichols.95}, we have
\begin{equation}
p(\theta_{mj} | \eta_{ms}) = \frac{1}{B(F \eta_{ms}, F(1-\eta_{ms}) )} \theta_{mj}^{ F \eta_{ms} -1} (1-\theta_{mj})^{ F(1-\eta_{ms})-1}. 
\end{equation}
Combine above two equations and drop the $m$ in notation, we have for an arbitrary marker
$$f(\theta_j, \eta_s) = \sum_{j=1}^K \left[ -\log B(F \eta_s, F(1-\eta_s) ) + (F \eta_s -1)~\log \theta_j + (F(1-\eta_s)-1)~\log (1-\theta_j)\right] p_{\cdot js}.$$

\begin{equation}
\begin{aligned}
\frac{d}{d \theta_j} f(\theta_j, \eta_s) &= \left[\frac{F \eta_s -1}{\theta_j} - \frac{F(1-\eta_s)-1}{(1-\theta_j)}\right] p_{\cdot js} 
\end{aligned}
\end{equation}
This suggest that we add $(F\eta_s -1) p_{\cdot js}$ to the top and $(F-2) p_{\cdot js}$ to the bottom of ~\eqref{eqn:tj} to estimate $\theta_j$.

\begin{equation}
\begin{aligned}
\frac{d}{d \eta_s} f(\theta_j, \eta_s) &=\sum_{j=1}^K \left[~-\frac{1}{B(F \eta_s, F(1-\eta_s) )} \frac{d}{d\eta_s} B(F \eta_s, F(1-\eta_s))  + F \log\frac{\theta_j}{1-\theta_j}~\right] p_{\cdot js} \\
&= - F \sum_{j=1}^K {p_{\cdot js}} ~\left[\Gamma(F \eta_s) - \Gamma(F(1-\eta_s))\right] +  F \sum_{j=1}^K\log\frac{\theta_j}{1-\theta_j} p_{\cdot js}
\end{aligned}
\end{equation}
where $\Gamma$ is a digamma function.
When $F > 1$, we use recurrence relation $\Gamma(x+1)  = 1/x + \Gamma(x)$ twice to get 
\begin{equation}
\begin{aligned}
\Gamma(F \eta_s) &= \Gamma(F \eta_s +2) - \frac{1}{F \eta_s + 1} - \frac{1}{F \eta_s}  \\
\Gamma(F (1- \eta_s)) &=  \Gamma(F (1-\eta_s)+2) - \frac{1}{F (1-\eta_s) + 1} - \frac{1}{F (1-\eta_s)} 
\end{aligned}
\end{equation}
Because $\eta \in [0,1]$, we may use  $\frac{1}{\exp(\Gamma(x))} = \frac{1}{x} + \frac{1}{2 x^2} + \frac{5}{4\cdot 3! x^3} + \frac{3}{2\cdot 4! \cdot x^4} + \frac{47}{48 \cdot 5! \cdot x^5}$ at $x=F\eta_s + 2$ and $x=F(1-\eta_s)+2$ to solve for $\eta_s$ numerically.  
When $F=1$, however, we may use the refection formula 
   $ \Gamma(1 - \eta_s) - \Gamma(\eta_s) = \pi\,\!\cot{ \left ( \pi \eta_s \right ) } $ to solve for $\eta_s$ analytically. 
   
The forward and backward probabilities of the ancillary HMM and other parameter estimates are simply special cases of the main HMM. 

\newpage

\setcounter{figure}{0}
\makeatletter 
\renewcommand{\thefigure}{S\@arabic\c@figure} 

\section{Supplementary Figures}
\vspace{1in}
\begin{figure}[h]
\centering
\includegraphics[width=7cm]{\dir/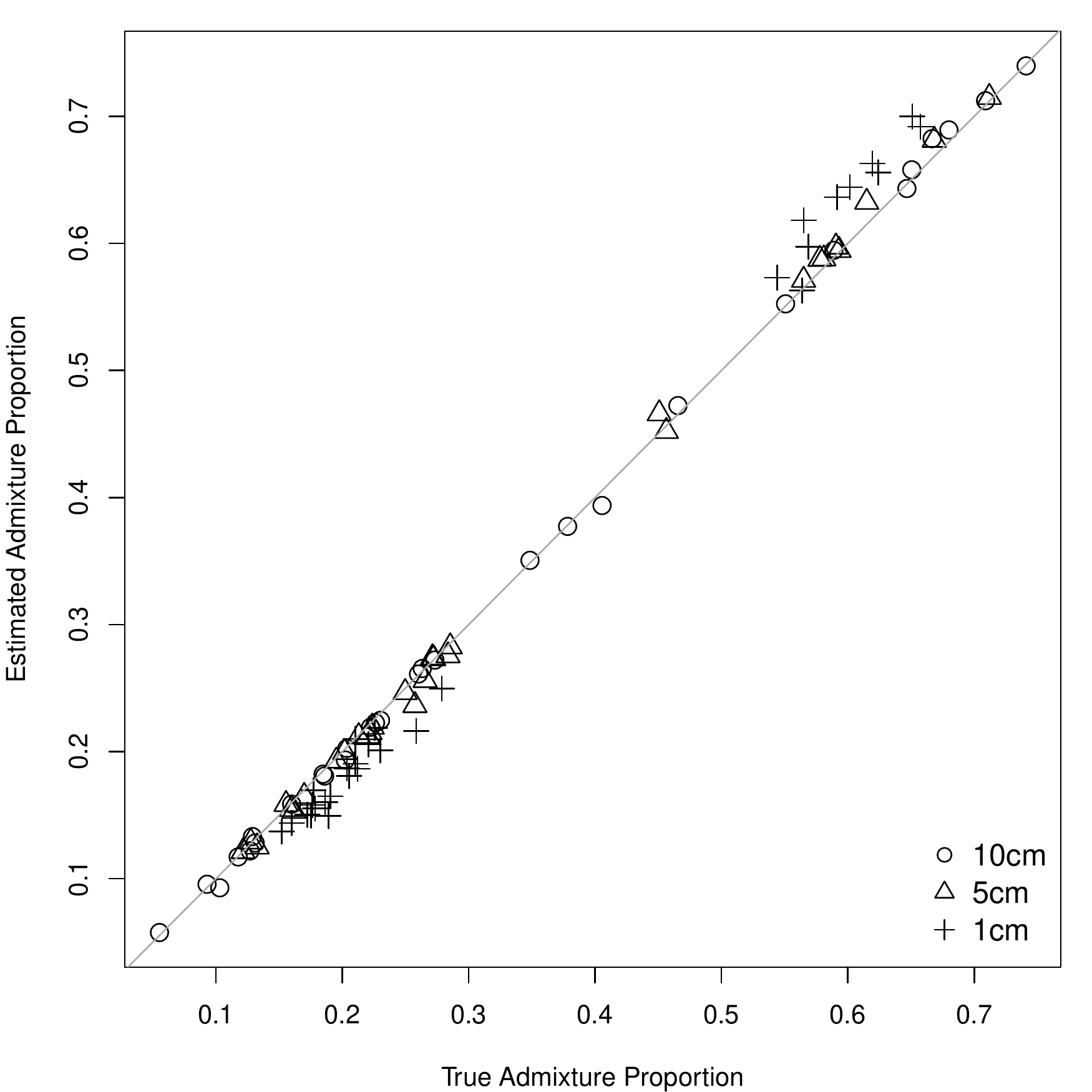}
\caption{Inference of admixture proportion.  The x-axis denotes the truth of admixture proportions, and the y-axis denotes inferred values. The gray line indicates $x=y$. For a three-way admixed individual there are three numbers (that sum to $1$) to denote admixture proportions. For admixture events that happened recently ($\g=10, 20$), the inference of admixture proportions is very accurate;  for remote admixture events ($\g=100$), our method slightly over-estimates large admixture proportions and slightly under-estimates the small ones.  }
\label{fig:admixture-proportion}
\end{figure}

\begin{figure}[h]
\centering
\scalebox{1}{
\includegraphics[width=12cm]{\dir/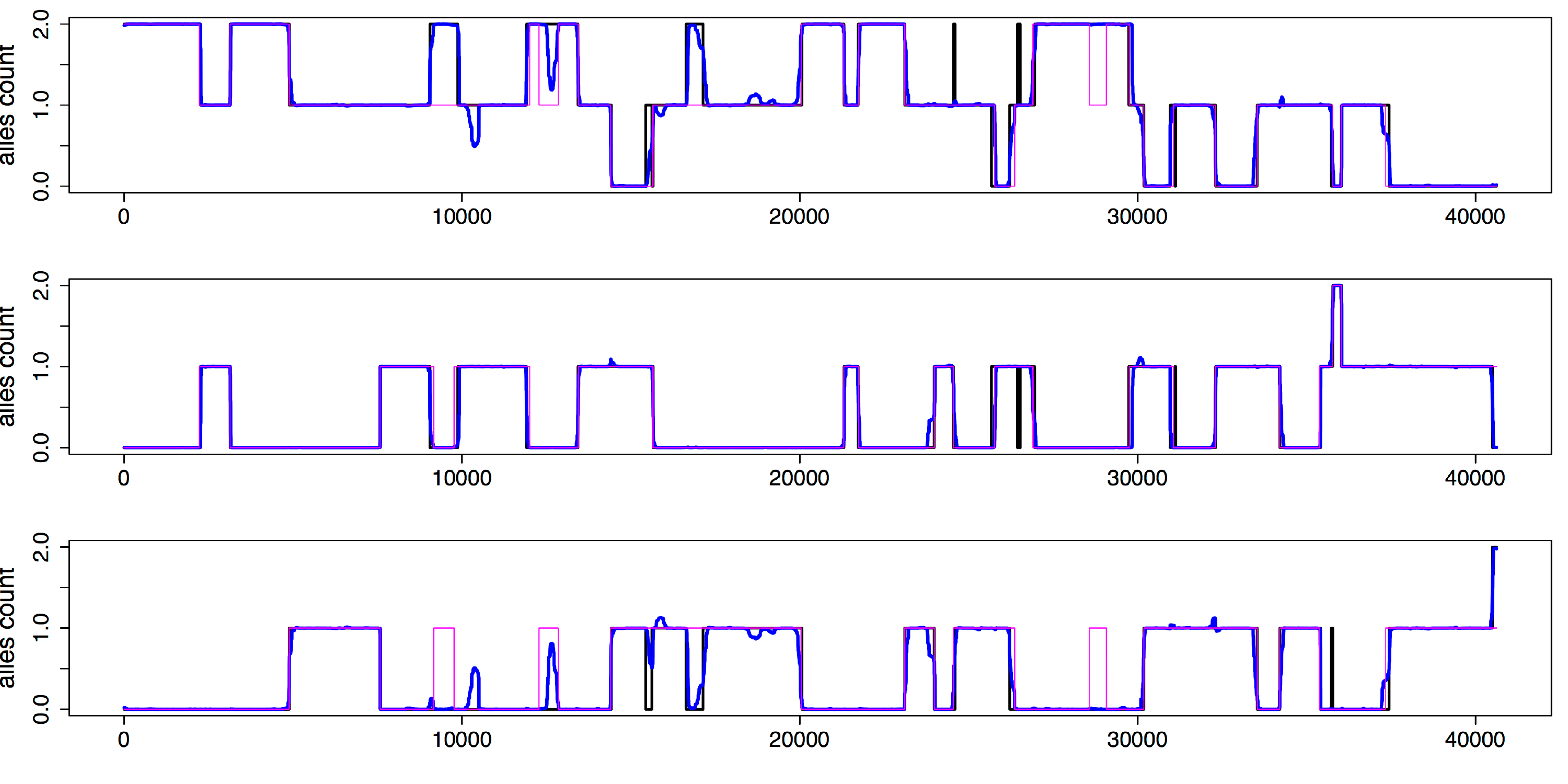}
}
\caption{Detailed comparison with LAMP-LD.  The plots shows the comparison for a typical simulated individual using $g=20$. Each panel denotes an ancestry, on which we plot the local ancestry of the actual (black line), our inference (blue line), and LAMP-LD inference (pink line). Compare to our method, LAMP-LD makes more mistakes on regions of a few hundred SNPs.}
\label{fig:lampld-detail}
\end{figure}

\begin{figure}[h]
\centering
(a) \includegraphics[width=12cm,height=6cm]{\dir/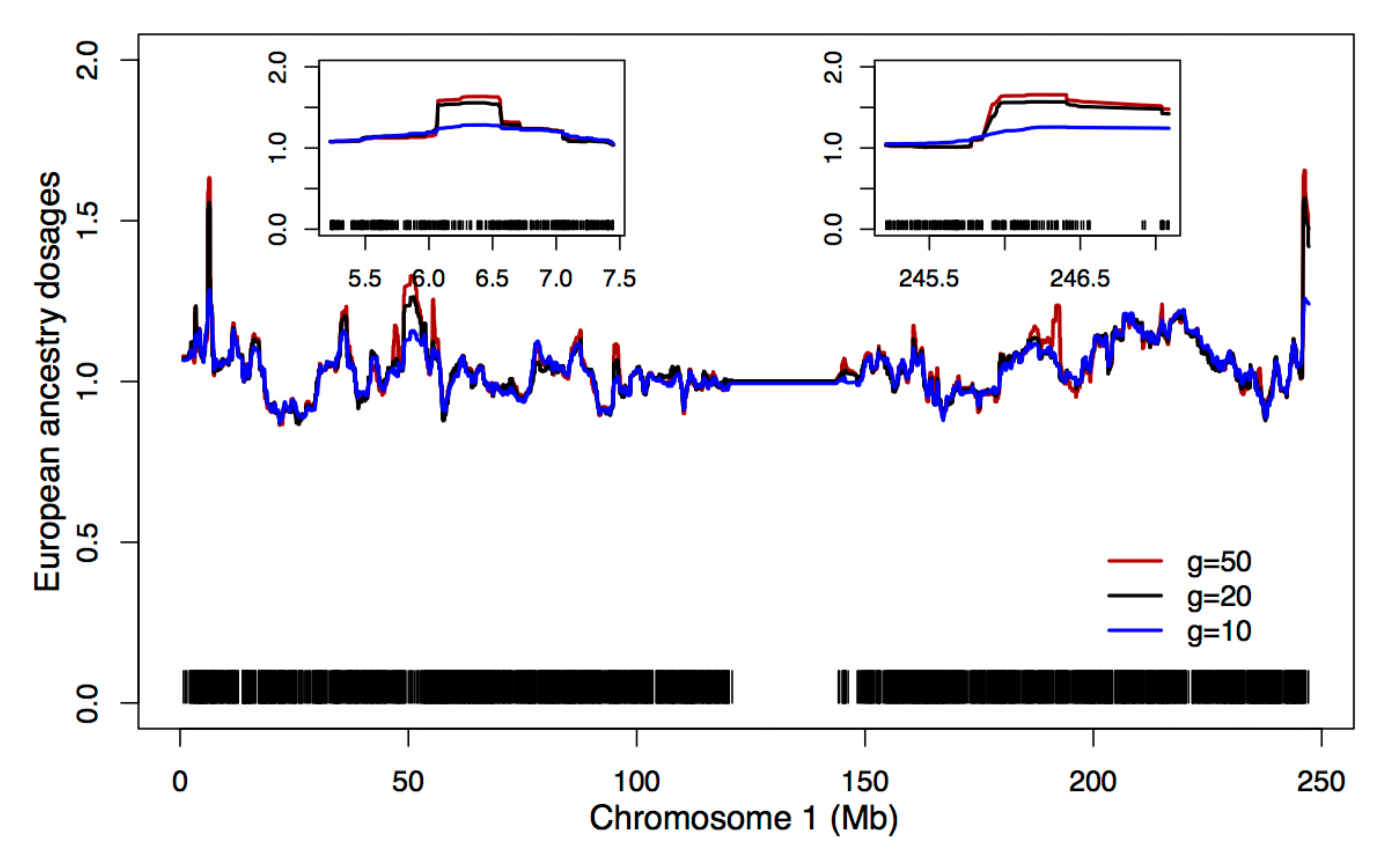}\\
(b) \includegraphics[width=12cm,height=6cm]{\dir/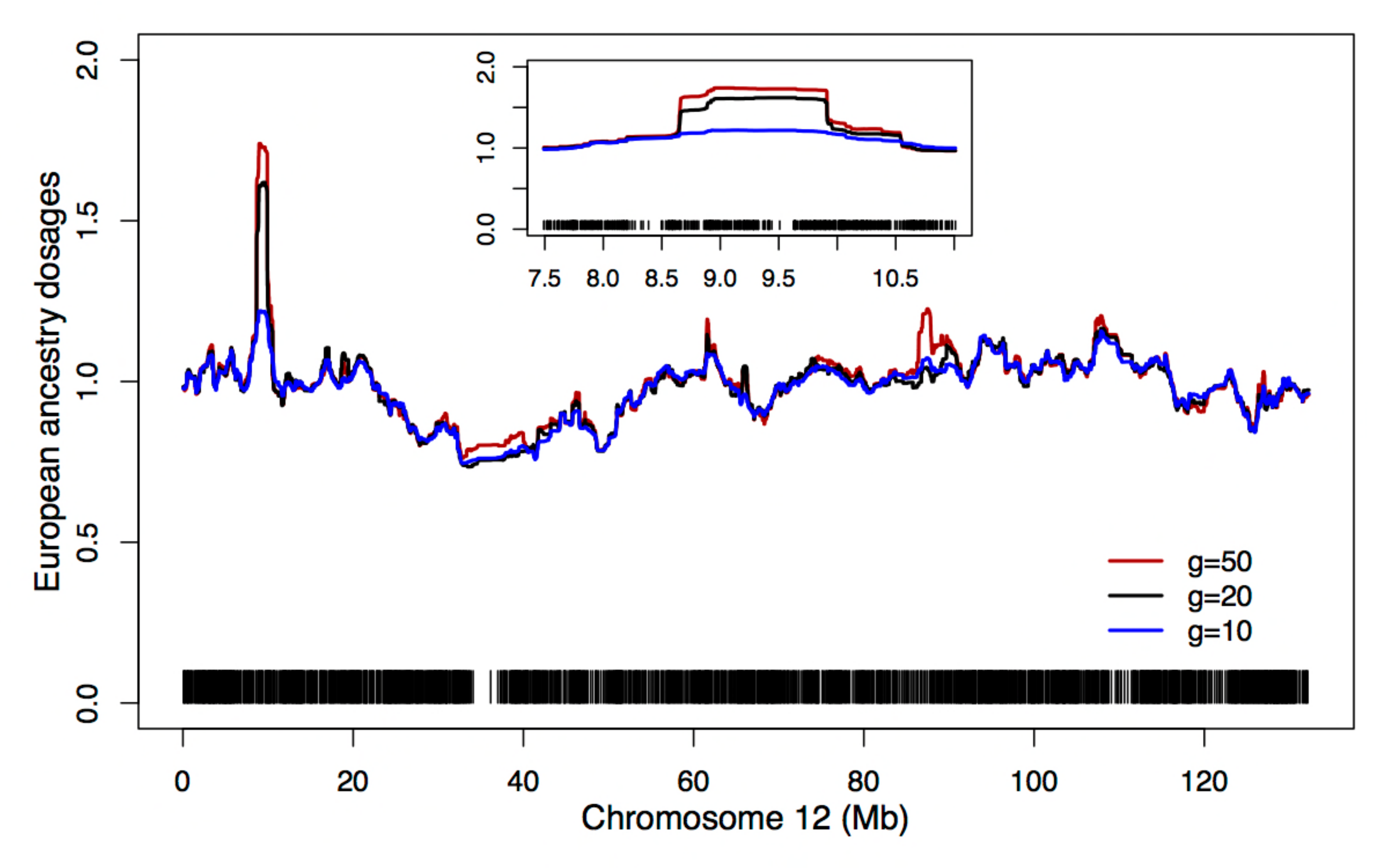}\\
(c) \includegraphics[width=12cm,height=6cm]{\dir/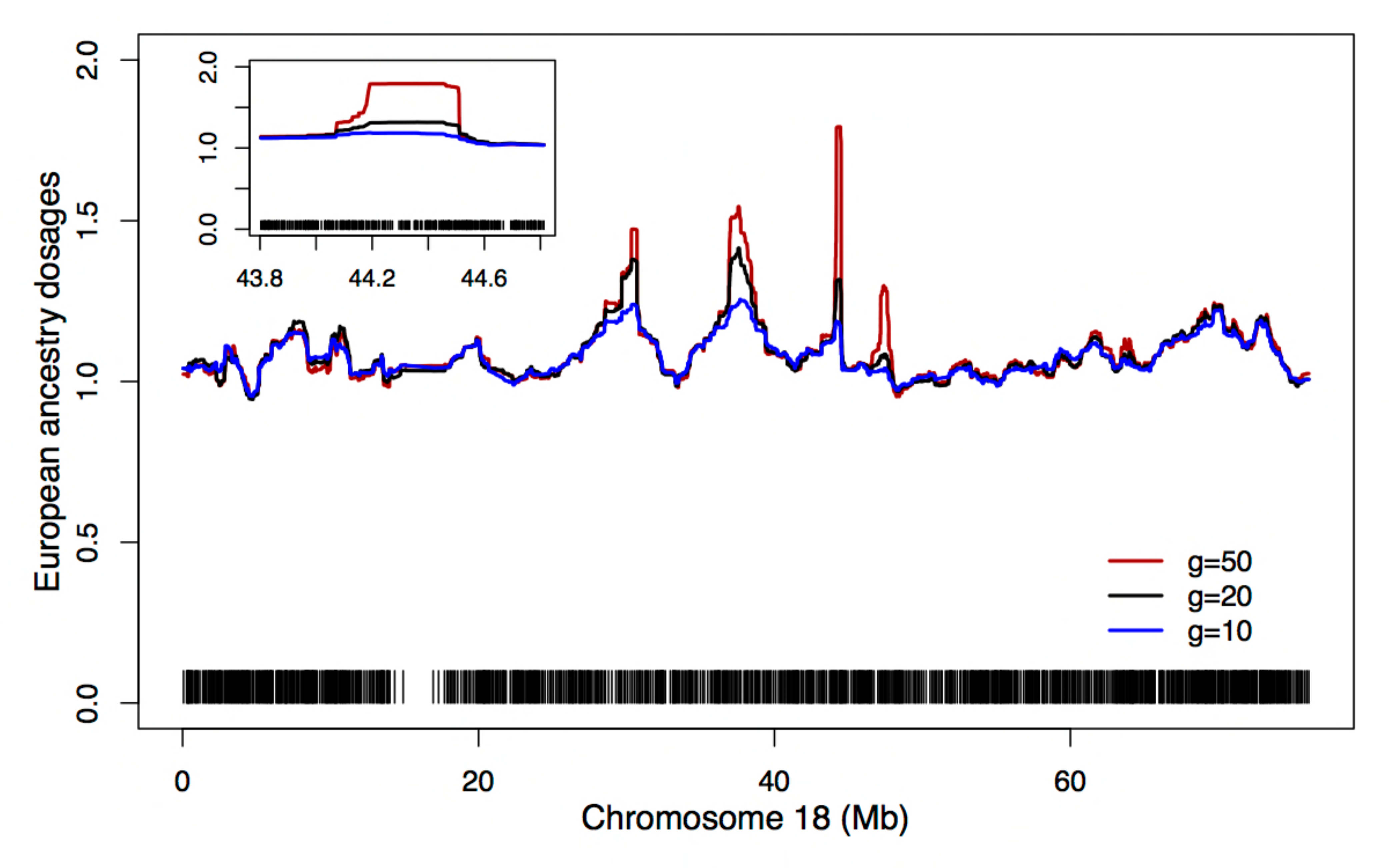}

\caption{Regions that exhibit excessive European ancestry in Mexicans.  The y-axis is the average European ancestry allele dosages (over $58$ Mexican samples). The $g$ in legends means $\g$ in main text. The inlets are close-ups of top peak(s) in each panel. The top panel contains two peaks, one is in 1p36.31 from $6.25$Mb to $6.55$Mb with average dosage of $1.63$, and the other is in 1q44 from 246.00Mb to 246.40Mb with average dosage of $1.65$; The middle panel contains one peak in 12p13.32 from $8.66$Mb to $9.92$Mb with average dosage of $1.72$; the bottom panel contains one peak in 18q21.1 from $44.18$Mb to $44.51$Mb with average dosage of $1.78$.}
\label{fig:enrich-all}
\end{figure}

\end{document}